\documentclass[aps,prresearch,twocolumn]{revtex4-2} 

\usepackage[]{graphicx}
\usepackage[dvipsnames]{xcolor}
\usepackage[colorlinks=true,citecolor=NavyBlue,linkcolor=RubineRed,urlcolor=Cerulean,pdfencoding=auto]{hyperref}
\usepackage[]{amsmath}
\usepackage[]{amssymb}
\usepackage{tabstackengine}
\usepackage{bbold}

\begin{document}
\title{Crosstalk analysis in single hole-spin qubits within highly anisotropic g-tensors }

\author{Yaser Hajati}
\email[]{yaser.hajati@uni-konstanz.de}
\affiliation{Department of Physics, University of Konstanz, D-78457 Konstanz, Germany}

\author{Irina Heinz}
\email[]{irina.heinz@uni-konstanz.de}
\affiliation{Department of Physics, University of Konstanz, D-78457 Konstanz, Germany}
\author{Guido Burkard}
\email[]{guido.burkard@uni-konstanz.de}
\affiliation{Department of Physics, University of Konstanz, D-78457 Konstanz, Germany}

\begin{abstract}
  Spin qubits based on valence band hole states are highly promising for quantum information processing due to their strong spin-orbit coupling and ultrafast operation speed. As these systems scale up, achieving high-fidelity single-qubit operations becomes essential. However, mitigating crosstalk effects from neighboring qubits in larger arrays, particularly for anisotropic qubits with strong spin-orbit coupling, presents a significant challenge. We investigate the impact of crosstalk on qubit fidelities during single-qubit operations and derive an analytical equation that serves as a synchronization condition to eliminate crosstalk in anisotropic media. Our analysis proposes optimized driving field conditions that can robustly synchronize Rabi oscillations and minimize crosstalk, showing a strong dependence on qubit anisotropy and the orientation of the external magnetic field. Taking experimental data into our analysis, we identify a set of parameter values that enable nearly crosstalk-free single-qubit gates, thereby paving the way for scalable quantum computing architectures.

\end{abstract}

\maketitle

\section{INTRODUCTION}

Spin qubits, using localized electron spins in quantum dots, are highly promising for quantum computing due to their ability to be coherently controlled using techniques such as electric-dipole-induced spin resonance \cite{Nowack,PioroLadrire,Golovach} and exchange-based gates \cite{Burkard,Brunner,zajac,yoneda,Watson}. However, in addition to challenges posed by charge noise and phonon-induced decoherence, electron spins are also subject to fluctuating nuclear spins, which can further complicate qubit stability and coherence \cite{Khaetskii,Bluhm}.
Valence-band hole spin qubits are emerging as a promising alternative for quantum computing due to several advantageous properties, including weak coupling between holes and residual nuclear spins, in combination with isotopically purified material \cite{Itoh}, a low effective mass, and the absence of a valley degeneracy in the valence band, which simplify device design \cite{Lodari}. Moreover, the strong spin-orbit coupling in hole states not only enables fast all-electric operation but also gives rise to highly anisotropic spin properties. This anisotropy is a distinct advantage, as it provides enhanced control over qubit manipulation and allows for greater precision in quantum gate operations\cite{Scappucci,Hendrickx,Lawrie,Jirovec,Hendrickx2,vanRiggelen, Borsoi}. In fact, the anisotropic nature of hole spin qubits, characterized by an anisotropic g-tensor, leads to a field-dependent magnetic response that deviates from a simple scalar g-factor \cite{Zhang,Jirovec2,Piot,Scherbl,Crippa}. The anisotropic g-tensor arising from the interplay between spin-orbit interaction, electric fields, and mechanical strain is significantly influenced by the symmetry-breaking effects of the nanostructured environment in double quantum dots (DQDs). This interaction further enhances the anisotropy, contributing to the non-uniformity and misalignment of the g-tensors within quantum dot arrays, which can be leveraged to improve control and precision in quantum operations \cite{Kloeffel,Kloeffel2,Nestoklon}.


Although these anisotropic properties can enhance certain aspects of qubit performance, they also introduce complexities, particularly in the form of crosstalk between qubits. The anisotropy exacerbates the coupling of hole spin qubits to charge noise \cite{Hendrickx3,yoneda}, heightening their sensitivity to variations in quantum dot confinement \cite{Jirovec2,Hendrickx2}. This increased sensitivity results in significant challenges for maintaining qubit fidelity, as crosstalk can lead to unintended interactions with neighboring qubits, impacting the overall performance and scalability of the qubit system. Despite these challenges, controlled anisotropy can be leveraged to create operational sweet spots where qubit control is optimized and decoherence is minimized \cite{Hendrickx3,Wang2021,Bosco2021,Wang2}. Thus, effectively addressing crosstalk is crucial for error prevention and achieving high-fidelity operation, which is essential for the successful scaling up of qubit systems \cite{Zajac2016,Ferdous2018,Irina,Irina2,Undseth}.

This paper specifically examines crosstalk in hole spin qubits, focusing on how the anisotropy of the g-tensor—due to strong spin-orbit coupling and the direction of the external magnetic field—affects interactions with neighboring qubits. Understanding and mitigating the impact of such anisotropy-induced crosstalk is essential for improving the fidelity and performance of qubit arrays \cite{Geyer,Lawrie,Hendrickx2,Hendrickx3,Hendrickx,Sarkar2023,Cifuentes,Wang2024}. Our analysis proposes optimized driving field conditions that robustly synchronize Rabi oscillations and minimize crosstalk, with a strong dependence on the anisotropy of the qubit and the orientation of the external magnetic field. By analyzing holes in germanium (Ge) and silicon (Si), we show that the effectiveness of the optimization varies significantly between materials, emphasizing the importance of the material-specific behavior of the g-tensor in reducing crosstalk and enhancing high-fidelity quantum operations.

The remainder of this paper is organized as follows. In Section~\ref{sec:stiment}, we introduce the effective Hamiltonian for a single spin qubit in anisotropic media, driven by electric-dipole spin resonance (EDSR). In Section~\ref{sec:stiment1}, we present the optimal parameters to minimize crosstalk errors in anisotropic media through simple synchronization schemes applicable to quantum algorithms, and outline the conditions to maximize the overall fidelity. In Section~\ref{sec:stimresults}, we analyze the relationship between crosstalk, the direction of the external magnetic field, and material anisotropy, leveraging experimental data for hole qubits in Ge and Si. Finally, the conclusions of our work are summarized in Section~\ref{sec:stimconc}.

\begin{figure}[!ht]
\centering
\includegraphics[width=0.8\linewidth]{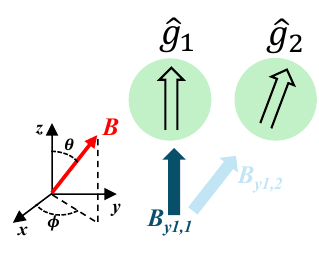} 
\caption{\label{fig:level1}
  Schematic representation of two hole spins in adjacent quantum dots. The influence of spin-orbit coupling results in the real and symmetric $g$-tensors \( \hat{g}_1 \) and \( \hat{g}_2 \). An external magnetic field \( \mathbf{B} \) with amplitude \( B \) and orientation defined by the angles \(\theta\) and \(\phi\) interacts with the spins through their respective \( g \)-tensors. The driving strength of qubit 1 is denoted as \( B_{y1,1} \), which, due to crosstalk, induces a driving effect on qubit 2 with strength \( B_{y1,2} \).
}
\end{figure}

%

\section{THEORETICAL MODEL }\label{sec:stiment}
We first examine a gate-defined quantum dot array operating in the (1, 1, ...) charge regime, as illustrated in Fig.~\ref{fig:level1}. Next, we consider a widely used phenomenological \(2 \times 2\) model Hamiltonian that describes a single hole in a spin-orbit-coupled system \cite{Burkard2,Scherbl,Frank}. The Hamiltonian is given by 

\begin{align}
 \label{eq:Hsys}
  H(t) =& \mu_B \mathbf{B} \cdot (\hat{g}_1 \mathbf{S}_1 + \hat{g}_2 \mathbf{S}_2) \nonumber \\
  &+ B_{y1,1} \sin(\omega t) S_{x1} + B_{y1,2} \sin(\omega t) S_{x2},
\end{align}
where \(\mathbf{S}_i = \boldsymbol{\sigma}_i /2\), \(\mu_B\) is the Bohr magneton and \(\hat{g}_i\) \((i=1,2)\) are the real-valued spin-orbit-affected g-tensors of the two holes, defined as:
\begin{equation}
\label{eq:Hsys1}
\hat{g}_i = 
\begin{pmatrix}
    g_{xx,i} & g_{xy,i} & g_{xz,i} \\
    g_{yx,i} & g_{yy,i} & g_{yz,i} \\
    g_{zx,i} & g_{zy,i} & g_{zz,i}
\end{pmatrix},
\end{equation}
for \(i = 1, 2\). Additionally, the Zeeman interaction with the external homogeneous magnetic field is expressed as \( \mathbf{B}= B \mathbf{b} \), where \(\mathbf{b} = \begin{pmatrix} \sin(\theta) \cos(\phi) & \sin(\theta) \sin(\phi) & \cos(\theta) \end{pmatrix}^T \) is the unit vector pointing along the direction of \(\mathbf{B}\), and $B=|\mathbf{B}|$, see Fig.~\ref{fig:level1}.

Leveraging the strong spin-orbit coupling, all-electrical spin control is achieved through EDSR \cite{Nowack,PioroLadrire,Golovach,Geyer}. To this end, rapid voltage pulses and microwave (MW) bursts, \( B_{y1,1} \sin(\omega t) \), are applied to qubit 1. Due to crosstalk, this leads to an induced microwave pulse \( B_{y1,2} \sin(\omega t) \) at the neighboring qubit, where the drive amplitude is given by \( B_{y1,2} = \alpha B_{y1,1} \).

We define the magnetic field-dependent g-factor 
\begin{eqnarray}
\label{eq:Hsys2}
& & \mathbf{g}_i^*(\theta, \phi)= \hat{g}_i \cdot \mathbf{b} =
\begin{pmatrix}
    A_i(\theta, \phi) \\
    B_i(\theta, \phi) \\
    C_i(\theta, \phi)
\end{pmatrix} \\
&=&
\begin{pmatrix}
    g_{xx,i} \sin(\theta)\cos(\phi)  + g_{xy,i} \sin(\theta) \sin(\phi) + g_{xz,i} \cos(\theta) \\
    g_{yx,i} \sin(\theta)\cos(\phi)  + g_{yy,i} \sin(\theta) \sin(\phi) + g_{yz,i} \cos(\theta) \\
    g_{zx,i} \sin(\theta)\cos(\phi) + g_{zy,i} \sin(\theta) \sin(\phi) + g_{zz,i} \cos(\theta)
\end{pmatrix}\nonumber
\end{eqnarray}
for \(i = 1, 2\), represents the components of the g-factor, which depend on the direction of the magnetic field, characterized by the angles \(\theta\) and \(\phi\). In the rotating frame, $\tilde{H}(t) = U^{\dagger} H U + i \dot{U}^{\dagger} U$, with $U_{\rm rot}(t) = \exp( -i \omega_0 t \mathbf{S}_z)$, we apply the rotating wave approximation (RWA), which yields both resonant and off-resonant Rabi terms. The off-resonant Rabi frequency of qubit 2 is then given by
\begin{align}
\tilde{\Omega} = \sqrt{\left( \Delta B\right)^2 + \left(\frac{\alpha B_{y1,1} |\zeta(\theta, \phi)|}{2}\right)^2},
\end{align}
where the Zeeman energy difference $\Delta B$ and driving amplitude coefficient due to the anisotropy $\zeta(\theta, \phi)$ are
\begin{widetext}
\begin{align}
\label{eq:Hsys 5}
 \Delta B=B \Delta g^*(\theta, \phi)
 = B\left(\sqrt{A_2(\theta, \phi)^2 + B_2(\theta, \phi)^2 + C_2(\theta, \phi)^2}
- \sqrt{A_1(\theta, \phi)^2 + B_1(\theta, \phi)^2 + C_1(\theta, \phi)^2}\right),\\
\zeta(\theta, \phi) = \frac{1}{\sqrt{A_2(\theta, \phi)^2 + B_2(\theta, \phi)^2}} 
\left( B_2(\theta, \phi) -i A_2(\theta, \phi)
\frac{C_2(\theta, \phi)}{\sqrt{A_2(\theta, \phi)^2 + B_2(\theta, \phi)^2 + C_2(\theta, \phi)^2}} \right),
\end{align}
\end{widetext}
with $\Delta g^*(\theta, \phi) = |\mathbf{g}_2^*(\theta, \phi)|-|\textbf{g}_1^*(\theta, \phi)|$.

\section{Synchronization to avoid crosstalk}\label{sec:stiment1}
To minimize the effects of crosstalk, we require that during the driving time \(\tau\) for spin rotations, neighboring spins perform full \(2\pi\) rotations, expressed as \(\tilde{\Omega} \tau = 2\pi k\), where \(k \in \mathbb{Z}\). Using the ratio of the induced driving amplitudes \(\alpha = \frac{B_{y_{1,2}}}{B_{y_{1,1}}}\), we find a condition for \(B_{y_{1,1}}\) in terms of \(\alpha\). The time to perform a \(\pi\) rotation on qubit~1 is given by \(\tau = \frac{\pi(2m+1)}{\Omega}\), where \(m \in \mathbb{Z}\), with the Rabi frequency for on-resonance driving of qubit 1 defined as 
\begin{align}
\label{eq:Hsys}
\Omega =& \frac{B_{y_{1,1}}}{2} \, |\mathbf{g}_1^*(\theta, \phi)|  \nonumber\\
& =\frac{B_{y_{1,1}}}{2}\sqrt{A_1(\theta, \phi)^2 + B_1(\theta, \phi)^2 + C_1(\theta, \phi)^2}.
\end{align}
This leads to the synchronization condition for eliminating crosstalk when driving a qubit in anisotropic media,
\begin{gather}
\label{eq:Hsys8}
B_{y_{1,1}} = \frac{2B \, \Delta g^*(\theta, \phi)}{\sqrt{\left( |\mathbf{g}_1^*(\theta, \phi)|  \frac{2k}{2m+1}\right)^2 - (\alpha |\zeta(\theta, \phi)|)^2}},
\end{gather}
for integers \(k\) and \(m\). For \( \alpha = 0 \), the remaining oscillation around the \( z \)-axis can be neglected, while for \( 0 < \alpha < 1 \), the fidelity reaches a maximum when the condition in Eq.~\eqref{eq:Hsys8} is fulfilled.
To evaluate the crosstalk of single-qubit gates on neighboring qubits, we calculate the fidelity \cite{Pedersen},
\begin{align}
    F = \frac{d + \left| \text{Tr}\left[ U_{\text{ideal}}^{\dagger} U_{\text{actual}} \right] \right|^2}{d(d+1)},
\end{align}
where \(d\) is the dimension of the Hilbert space, \(U_{\text{ideal}}\) is the desired qubit operation—which, in the case of crosstalk, would be \(\mathbb{1}\) for the neighboring qubits—and \(U_{\text{actual}}\) is the actual operation, containing unwanted off-resonant Rabi oscillations with a detuned Rabi frequency,
\begin{align}
    \tilde{\Omega} = \sqrt{\Omega^2 + \delta \omega_z^2}. \label{Rabifreq}
\end{align}
Here, $\Omega$ denotes the resonant Rabi frequency and $\delta \omega_z$ represents the detuning between the driving and resonance frequencies. Note that to describe the time evolution of a quantum state \(\vert \psi(t) \rangle\) in a frame rotating with its own qubit frequency, taking into account both the rotating transformation and the RWA Hamiltonian, we have
\begin{align}
\vert \psi(t) \rangle_{\text{lab}} &= U_{\text{actual}} \vert \psi(0) \rangle =  U_{\text{z}}(t) U_{\text{RWA}}(t)  \vert \psi(0) \rangle \nonumber\\ &=e^{-i  \Delta B t \mathbf{S}_z} e^{-i H_{\text{RWA}} t} \vert \psi(0) \rangle,
\label{eq:time_evolution}
\end{align}
where \(U_{\text{actual}} = U_{\text{z}}(t) U_{\text{RWA}}(t)\) represents the complete time evolution operator in the rotating frame. Here, \(U_{\text{z}}(t) =e^{-i  \Delta B t \mathbf{S}_z}\) is a $z$-rotation with the qubit frequency and \(U_{\text{RWA}} = e^{-i H_{\text{RWA}} t}\) describes the time evolution in the rotating frame under the RWA Hamiltonian.


\section{Results}\label{sec:stimresults}
In the following, we focus on the performance of \(y\)-rotations (Y-gate) performed on qubit 1 for single-qubit operations, and their corresponding crosstalk on the nearest-neighbor qubit 2, which manifests as an unwanted microwave (MW) driving field on the neighboring qubit. In the case of EDSR, this crosstalk can be capacitively induced \cite{Kelly,Cayao,Neyens2019} by the actual driving field \(B_{y1,1}\) (\(B_{y1,2}\)) applied on the corresponding gate \cite{Nowack}. All operations are performed in the presence of strong spin-orbit coupling, which plays a crucial role in the mechanism of the spin control and crosstalk effects.

\begin{figure*}[ht!]
\centering
\includegraphics[width=0.95\linewidth]{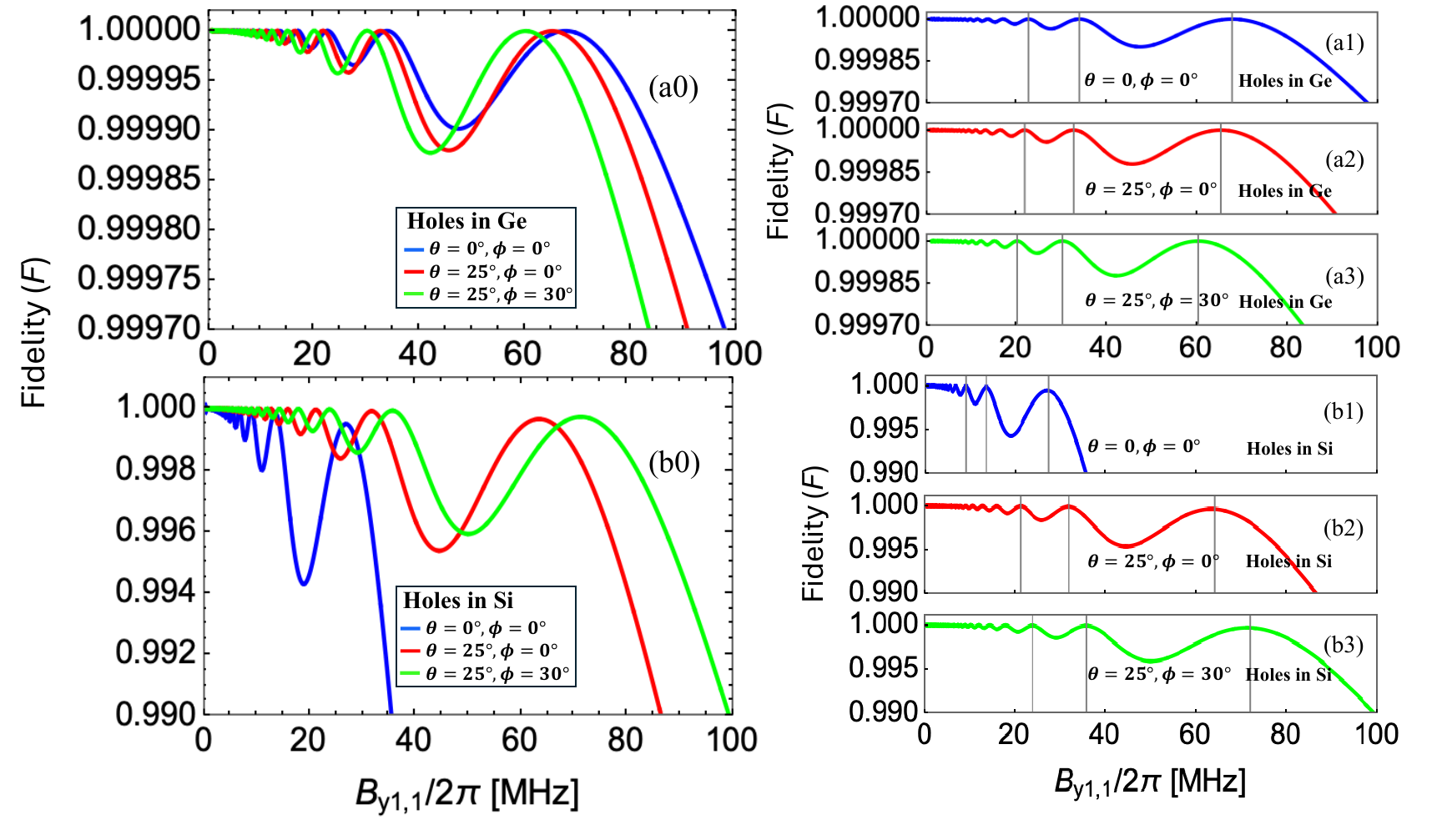} 
\caption{\label{fig:level2}
The fidelity $F$ of the adjacent qubit 2, i.e., probability of remaining in its state, while driving qubit 1, assuming $\alpha=0.4$, is influenced by the driving strength $B_{y1,1}$ and the direction of the magnetic field defined by the angles $\theta$ and $\phi$. This is illustrated for (a0) holes in Ge and (b0) holes in Si. Specifically, for holes in Ge, we have (a1) with $\theta=\phi=0^\circ$ (a2) for $\theta=25^\circ$ and $\phi=0^\circ$, and (a3) for $\theta=25^\circ$ and $\phi=30^\circ$. For holes in Si, the corresponding configurations are (b1) with $\theta=\phi=0^\circ$, (b2) for $\theta=25^\circ$ and $\phi=0^\circ$, and (b3) for $\theta=25^\circ$ and $\phi=30^\circ$. Here, the external magnetic field \(B\) is 1020 MHz. 
}
\end{figure*}

After deriving Eq.~\eqref{eq:Hsys8} to mitigate crosstalk, we plot the fidelity in Fig.~\ref{fig:level2} as a function of the drive amplitude \( B_{y_{1,1}} \) for two anisotropic materials and different magnetic field directions to analyze how the fidelity varies with the drive strength. To ensure that our analysis is grounded in realistic scenarios, we incorporate experimental data for the \( g \)-tensor in anisotropic media, specifically for holes in Si and holes in Ge, which are among the most promising materials for spin-based quantum computing. As an example, the full \(g\)-tensors for holes in a pair of adjacent quantum dots in Si were found to be \cite{Geyer}
\begin{align}
\label{eq:Hsys1}
\hat{g}_1 &= \begin{pmatrix} 2.31 & 0.50 & -0.06 \\ 0.50 & 2.00 & 0.06 \\ -0.06 & 0.06 & 1.50 \end{pmatrix}, 
\\
\hat{g}_2 &= \begin{pmatrix} 1.86 & -0.57 & 0.09 \\ -0.57 & 2.76 & -0.01 \\ 0.09 & -0.01 & 1.46 \end{pmatrix}.
\end{align}
Another experiment reveals the full \(g\)-tensors for holes in a Ge double quantum dot as \cite{Hendrickx3}
\begin{align}
\label{eq:Hsysa10}
\hat{g}_1 &= \begin{pmatrix} 
    0.082 & 0.018 & -0.494 \\ 
    0.018 & 0.394 & -0.020 \\ 
    -0.494 & -0.020 & 11.233 
\end{pmatrix}, \\[10pt]
\hat{g}_2 &= \begin{pmatrix} 
    0.0643 & 0.006 & -0.214 \\ 
    0.006 & 0.358 & 0.038 \\ 
    -0.214 & 0.038 & 10.945 
\end{pmatrix}.
\end{align}

\begin{figure}[t]
\centering
\includegraphics[width=\linewidth]{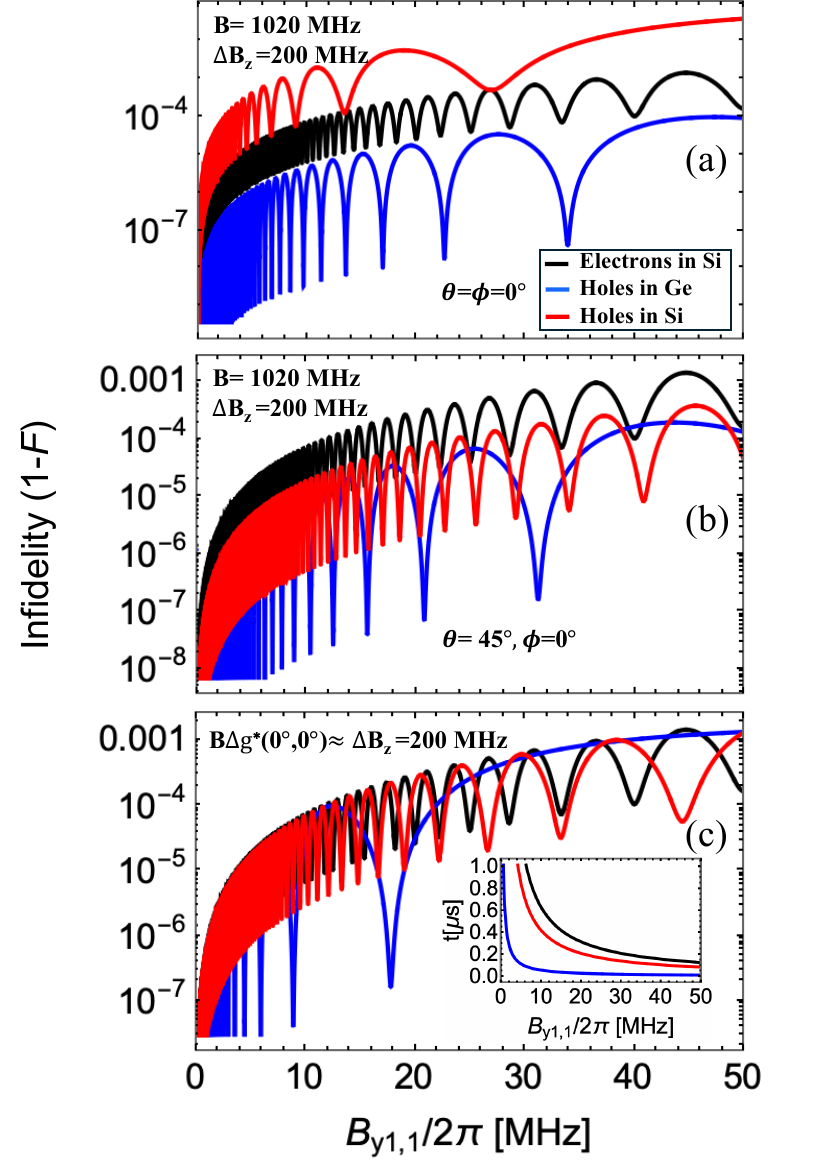} 
\caption{\label{fig:level3}
  The infidelity 1-$F$ of qubit 2 as a function of driving strength $B_{y1,1}$ for different material scenarios, including electrons in silicon with $\Delta B_z = 200 \, \text{MHz}$ and holes in Ge and Si. A fixed magnetic field of $B = 1020 \, \text{MHz}$ is analyzed for (a) $\theta = \phi = 0^\circ$ where \( \Delta g^*(0^\circ,0^\circ)\) for holes in Ge and Si are 0.75 and 0.04, respectively and (b) $\theta = 45^\circ$, $\phi = 0^\circ$ where \( \Delta g^*(45^\circ,0^\circ)\) for Ge and Si are 0.49 and 0.37, respectively. In (c), we maintain a fixed \( B \Delta g^*(0^\circ,0^\circ)\) for holes in Ge and Si, ensuring that this matches \( \Delta B_z = 200 \, \text{MHz} \), which is typically used for electrons in silicon. Consequently, we obtain \( B = 268 \, \text{MHz} \) for holes in Ge and \( B = 5051 \, \text{MHz} \) for holes in Si, respectively. The inset in (c) shows the gate time versus drive strength $B_{y1,1}$ for electrons in Si and holes in Ge and Si.}
\end{figure}

For these numerical examples, we assume a crosstalk coupling of \(\alpha = 0.4\) \cite{Irina,Irina2} and a magnetic field of \( B = 1020 \, \text{MHz} \) (except in Fig.~\ref{fig:level3}(c)). We vary the nearest-neighbor qubit crosstalk field \( B_{y1,2} \), as well as the term \( B \Delta g^*(\theta, \phi) = B (|\mathbf{g}_2^*(\theta, \phi)| - |\mathbf{g}_1^*(\theta, \phi)|) \), while investigating idle qubit 2. In Fig.~\ref{fig:level2}(a0), for holes in Ge, we observe that increasing the drive amplitude \( B_{y1,1} \) leads to a decrease in the oscillation frequency of fidelity. Altering the direction of the magnetic field modifies the g-factor, denoted as $\mathbf{g}_i^*(\theta, \phi)$ in Eq.~\eqref{eq:Hsys2}, which results in fidelity modulation and affects qubit efficiency. In Figs.~\ref{fig:level2}(a1)--(a3), we illustrate that, for holes in Ge, fidelity maxima occur when the synchronization condition (Eq.~\eqref{eq:Hsys8}) is met, corresponding to drive amplitudes where crosstalk is fully suppressed. This suppression is marked by the vertical lines in Figs.~\ref{fig:level2}(a1)--(a3) and (b1)--(b3), with the three lines, from right to left, representing \( k = 1 \), \( k = 2 \), and \( k = 3 \) for \( m = 0 \). For example, in Figs.~\ref{fig:level2}(a1)--(a3) for holes in Ge, at \( \theta = \phi = 0^\circ \), crosstalk occurs at 67.75 MHz; this shifts to 65.25 MHz for \( \theta = 25^\circ, \phi = 0^\circ \), and drops to 60.23 MHz for \( \theta = 25^\circ, \phi = 30^\circ \), all corresponding to \( k = 1 \), the first peak on the right. These observations illustrate how adjusting the magnetic field angles can shift the crosstalk location, allowing for precise control over its suppression. In Fig.~\ref{fig:level2}(b0), we plot the fidelity as a function of drive strength for holes in Si spin qubits at various angles of the external magnetic field. Compared to the fidelity pattern in Fig.~\ref{fig:level2}(a0), we find significant changes, where fidelity maxima, corresponding to the absence of crosstalk, occur at different driving amplitudes. This variation is further demonstrated in Figs.~\ref{fig:level2}(b1)--(b3), where the shift in drive strength necessary for crosstalk suppression is evident for each angle. For instance, in Figs.~\ref{fig:level2}(b1)--(b3) for holes in Si, at \( \theta =\phi = 0^\circ \), crosstalk occurs at 27.18 MHz; for \( \theta = 25^\circ, \phi = 0^\circ \), it shifts to 63.02 MHz, and for \( \theta = 25^\circ, \phi = 30^\circ \), it is 71.89 MHz, all corresponding to \( k = 1 \), the first peak on the right. By comparing Figs.~\ref{fig:level2}(a0) and ~\ref{fig:level2}(b0), it is evident that increasing the magnetic field angle from zero (\( \theta = \phi = 0^\circ \)) causes the fidelity peak associated with crosstalk elimination to shift in opposite directions with respect to the drive amplitude. This shift underscores the material and magnetic field direction's critical role in achieving the highest fidelity, which is essential for optimal qubit operation. Note also that synchronization for second or further neighbors, as demonstrated in \cite{Irina}, can be achieved to minimize crosstalk on multiple qubits. For hole spin qubits, the magnetic field direction and the anisotropy of second or further neighbors serve as additional tuning parameters to facilitate synchronization.

Figs.~\ref{fig:level3}(a) and \ref{fig:level3}(b) illustrate the infidelity \(1-F\) of qubit 2 as a function of the driving strength \(B_{y1,1}\) for various materials, including electrons in silicon with \(\Delta B_z = 200 \, \text{MHz}\) exhibiting isotropic behavior, as well as holes in Ge and Si that display anisotropic characteristics. We analyze a fixed magnetic field strength of \(B = 1020 \, \text{MHz}\) for holes in Ge and Si while varying \(\Delta g^*(\theta, \phi)\) through magnetic field direction for \(\theta = \phi = 0^\circ\) in (a) and \(\theta = 45^\circ\), \(\phi = 0^\circ\) in (b). It is observed that holes in Si, for a magnetic field in the $z$-direction (\(\theta = 0^\circ\)), show lower fidelity compared to electrons in Si which is due to a very small \(\Delta g^* (0^\circ,0^\circ)= 0.04\). However, changing the magnetic field direction to \(\theta = 45^\circ\) and \(\phi = 0^\circ\), which leads to \(\Delta g^* (45^\circ,0^\circ)= 0.37\), results in better fidelity for holes in Si, as can be seen in Fig.~\ref{fig:level3}(b). In Figs.~\ref{fig:level3}(a) and ~\ref{fig:level3}(b), holes in Ge exhibit very high fidelity compared to holes and electrons in Si, attributed to their significantly larger \(\Delta g^*(\theta,\phi)\), which is \(0.75\) at \(\theta = \phi = 0^\circ\) in Fig.~\ref{fig:level3}(a), and \(0.49\) at \(\theta = 45^\circ\) and \(\phi = 0^\circ\) in Fig.~\ref{fig:level3}(b).

In the case of electrons in Si with isotropic nature, different Zeeman fields are often applied to the left and right spin qubits, typically choosing \( \Delta B_z = 200 \, \text{MHz} \) \cite{zajac,Irina,Irina2}. For holes in anisotropic materials, \( B \Delta g^*(\theta, \phi) \) plays a similar role to \( \Delta B_z \) for electrons in silicon. This Zeeman splitting term, after applying the RWA, highlights how isotropic and anisotropic systems react differently to external magnetic fields, impacting spin qubit behavior. Hence, in Fig.~\ref{fig:level3}(c), we selected different magnetic field strengths $B$ for holes in Ge and Si to achieve the same \(\Delta B_z\) of \(200 \, \text{MHz}\) for electrons in Si \cite{Irina}, while maintaining the magnetic field in the $z$-direction (\(\theta = 0^\circ\)). The corresponding \(\Delta g^*(0^\circ, 0^\circ)\) values for Ge and Si are \(0.04\) and \(0.75\), respectively. By choosing \(B = 268 \, \text{MHz}\) and \(5051 \, \text{MHz}\) for holes in Ge and Si, respectively,  we ensure that \(B \Delta g^*(0^\circ, 0^\circ) = 200 \, \text{MHz}\), aligning with \(\Delta B_z\) for electrons in Si. It is evident that even when \(B \Delta g^*(0^\circ, 0^\circ)\) matches \(\Delta B_z\), holes in Ge with a larger \(  |\mathbf{g}^*_{1}(0^\circ, 0^\circ)|\) and consequently higher Rabi frequency exhibit an improved fidelity, as shown by the blue curve in Fig.~\ref{fig:level3}(c). Moreover, by adjusting the amplitude of \(B\), we can either confine or broaden the oscillations of the fidelity. For instance, with \(B = 268 \, \text{MHz}\) for holes in Ge—significantly lower than \(B = 1020 \, \text{MHz}\) in Fig.~\ref{fig:level3}(a)—we can effectively confine the fidelity oscillations at very low drive amplitudes. In contrast, increasing the amplitude of $B$ for holes in Si to \(5051 \, \text{MHz}\) relative to \(1020 \, \text{MHz}\) in Fig.~\ref{fig:level3}(a) results in broader oscillations in fidelity.

In the inset of Fig.~\ref{fig:level3}(c), we plot the gate time versus drive strength $B_{y1,1}$ for electrons in Si and holes in Ge and Si. The gate time is defined as \( \tau = \frac{\pi (2m+1)}{\Omega} \), where \( \Omega \) is the Rabi frequency of the first qubit. When the magnetic field is aligned along the \( z \)-direction (\( \theta = \phi = 0^\circ\)), the gate time for holes can be expressed as \( \tau = \frac{2\pi (2m+1)}{\sqrt{g_{xz1}^2 + g_{yz1}^2 + g_{zz1}^2}B_{y1,1}} \). Specifically, for the above example for holes in Si, this results in \( \tau = \frac{2\pi (2m+1)}{1.46 B_{y_{1,1}}} \), while for holes in Ge, it is \( \tau = \frac{2\pi (2m+1)}{11.24 B_{y_{1,1}}} \), as derived from the data for $\hat{g}_{1}$ in Eqs.~\eqref{eq:Hsys1} and ~\eqref{eq:Hsysa10}. These values are shorter than the gate time \( \tau = \frac{2\pi (2m+1)}{B_{y1,1}} \) for electrons in Si. It should be noted that, although \( B \Delta g^*(\theta, \phi) \) remains constant across all materials, the gate time is influenced by the anisotropy of the material, being reduced for anisotropic cases. The reduction in gate times significantly improves the fidelity for holes, which is due to the anisotropic nature of the holes, particularly those in Ge. This leads to a higher Rabi frequency and shorter gate time for Ge holes compared to both Si holes and electrons. We conclude that for both cases \(B \Delta g^*(\theta, \phi) \gg \Delta B_z\) as shown in Fig.~\ref{fig:level3}(b), and \(B \Delta g^*(\theta, \phi) = \Delta B_z\) in Fig. ~\ref{fig:level3}(c), materials with a larger \(\Delta g^*(\theta, \phi)\) and \(| \mathbf{g}^*_{1}(\theta, \phi)|\) consistently exhibit better fidelity \cite{Lawrie}. Furthermore, it is important to highlight that in all cases, the position of the crosstalk is significantly affected by the orientation of the magnetic field.

\begin{figure*}[ht!]
\centering
\includegraphics[width=1\linewidth]{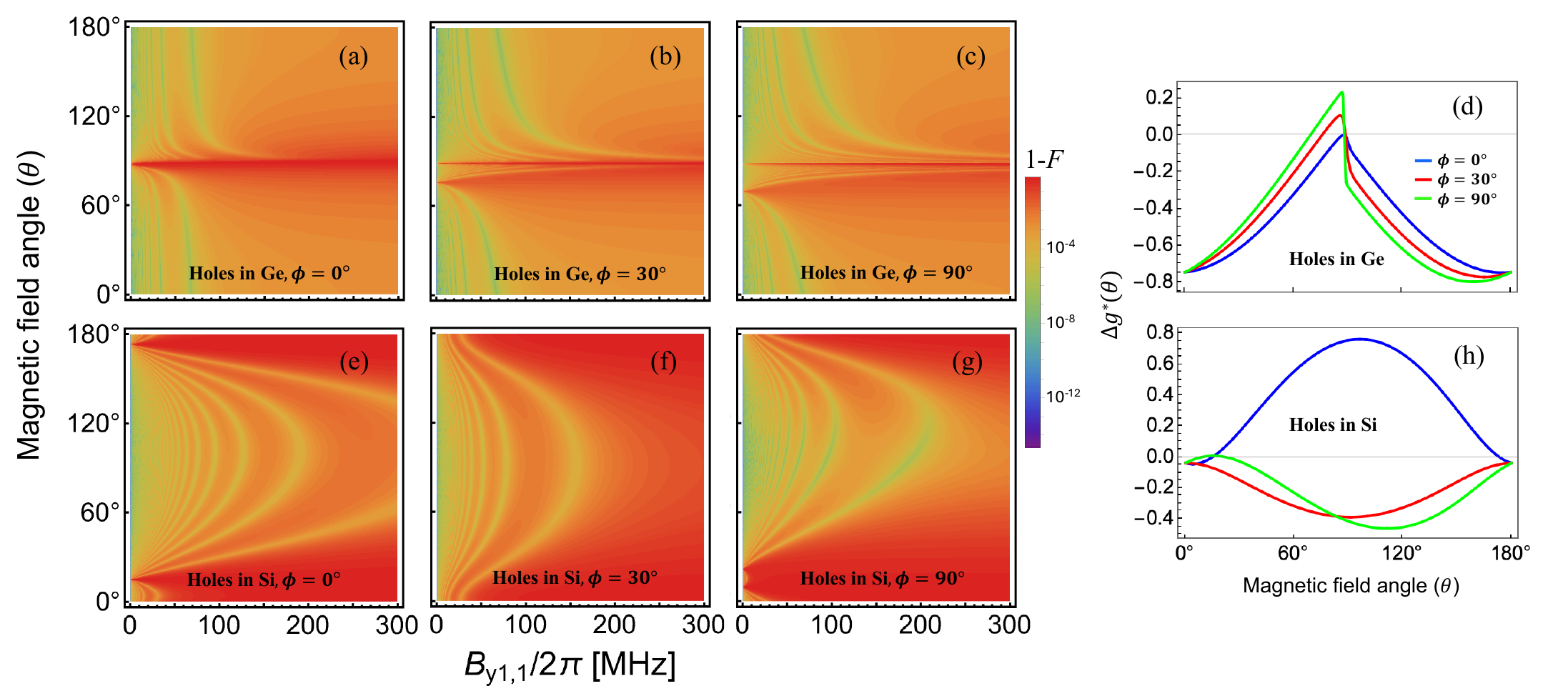}
\caption{The infidelity \(1-F\) of qubit 2 as a function of driving strength \(B_{y1,1}\) and magnetic field angle \(\theta\) for different values of the angle \(\phi\): (a)-(c) show results for holes in Ge and (e)-(g) for holes in Si. \(\Delta g^*(\theta)\) as a function of \(\theta\) for different values of \(\phi\) is plotted in (d) for holes in Ge and (h) for holes in Si.}
\label{fig:level4}
\end{figure*}

In Figs.~\ref{fig:level4}(a)-(c), we illustrate the infidelity \(1-F\) of qubit 2 as a function of the driving strength \(B_{y_{1,1}}\) and the magnetic field angle \(\theta\) for various values of \(\phi\) for holes in Ge. In Fig.~\ref{fig:level4}(a) with \(\phi=0^\circ\), the fidelity exhibits an oscillatory behavior as a function of the drive strength, except at one angle (\(\theta=90^\circ\)) where a notable decline occurs. Modifying \(\phi\) to \(30^\circ\) and \(90^\circ\) in Figs.~\ref{fig:level4}(b) and ~\ref{fig:level4}(c) reveals a similar pattern, introducing one additional point of reduced fidelity. 
To clarify this phenomenon, we plot \(\Delta g^*(\theta, \phi)\) in Eq.~\eqref{eq:Hsys 5} as a function of \(\theta\) for fixed \(\phi\) in Fig.~\ref{fig:level4}(d). In particular, for \(\phi=0^\circ\), \(\Delta g^*(\theta)\) goes to zero at \(\theta=87.71^\circ\), corresponding to the lowest fidelity observed in Fig.~\ref{fig:level4}(a). Furthermore, at \(\phi = 30^\circ\) (\(\phi = 90^\circ\)), we observe fidelity drops at two points corresponding to \(\theta = 75.79^\circ\) (\(\theta = 69.31^\circ\)) and \(\theta = 88.74^\circ\) (\(\theta = 88.11^\circ\)), which align with the zeros of \(\Delta g^*(\theta)\) seen in the red (blue) curve in Figs.~\ref{fig:level4}(b) and ~\ref{fig:level4}(c). When \(\Delta g^*(\theta)\) reaches zero, it implies that the \(| \mathbf{g}^*(\theta)|\) for both dots are identical. In this case, when there is no driving (\(B_{y1,1}=0\)), we observe the lowest fidelity, but applying a drive can shift these low-fidelity points to broader angles.

In Figs.~\ref{fig:level4}(e)-(g), we note similar behavior for holes in Si as compared to Ge. In these figures at \(\phi=0^\circ\) (\(\phi\) = \(90^\circ\)), the fidelity drop at two points correspond to \( \theta = 14.96^\circ \) (\( \theta = 9.40^\circ \)) and \( \theta = 173.30^\circ \)(\( \theta = 21.50^\circ \)), which correspond to the zeros of \(\Delta g^*(\theta)\) as can be seen in the blue (green) curve in Fig.~\ref{fig:level4}(h). By increasing \(\phi\) to \(30^\circ\) for holes in Si, we observe that the fidelity does not get worse at \(B_{y1,1}=0\) for \(0^\circ<\theta<180^\circ\), as confirmed by the red curve in Fig.~\ref{fig:level4}(h), where \(\Delta g^*(\theta)\) never reaches zero. Note also that the infidelity shows an oscillatory behavior with increasing drive amplitude and reaches a maximum corresponding to the elimination of crosstalk; this pattern changes for different angles. A comparison of the fidelity plots in Fig.~\ref{fig:level4} shows that certain angles of the external magnetic field, which vary depending on the material, result in the lowest fidelity. Avoiding these angles is important to optimize qubit performance. Additionally, the magnetic field direction affects not only the fidelity but also the crosstalk, as material-specific magnetic properties influence qubit interactions. Therefore, avoiding angles where \(\Delta g^*(\theta, \phi)\) approaches zero is essential to minimize crosstalk and ensure high-fidelity operation.

\section{Conclusions}\label{sec:stimconc}

In summary, we have analyzed crosstalk in spin qubits based on valence band hole states, which are highly promising candidates for quantum information processing due to their strong spin-orbit coupling and ultrafast operation speeds. As these systems scale up, the achievement of high-fidelity single-qubit operations becomes essential. However, mitigating crosstalk effects from neighboring qubits in larger arrays--particularly for anisotropic qubits with strong spin-orbit coupling--presents a significant challenge. We specifically investigate the impact of crosstalk on qubit fidelities during single-qubit operations, focusing on two realistic examples comprising holes in Ge and holes in Si. To address these challenges, we derive an analytical expression for the driving strength that serves as a synchronization condition to eliminate crosstalk in anisotropic media (Eq.~\eqref{eq:Hsys8}). Our analysis proposes optimized driving-field conditions that can robustly synchronize Rabi oscillations and minimize crosstalk, demonstrating a strong dependence on qubit anisotropy and the orientation of the external magnetic field. By incorporating experimental data into our analysis—specifically for holes in Ge and holes in Si, which are among the most promising spin qubit materials for quantum computing—we identify a set of parameter values that facilitate nearly crosstalk-free single-qubit gates. This research ultimately paves the way for scalable quantum computing architectures, significantly enhancing the feasibility of deploying these advanced qubit systems in practical applications.


\begin{acknowledgments}
  We acknowledge financial support from the IGNITE project of European Union’s Horizon Europe Framework Programme under grant agreement No.~101069515, and Army Research office (ARO) under grant W911NF-23-1-0104.
\end{acknowledgments}

\bibliography{refs.bib}

\begin{thebibliography}{46}%
\makeatletter
\providecommand \@ifxundefined [1]{%
 \@ifx{#1\undefined}
}%
\providecommand \@ifnum [1]{%
 \ifnum #1\expandafter \@firstoftwo
 \else \expandafter \@secondoftwo
 \fi
}%
\providecommand \@ifx [1]{%
 \ifx #1\expandafter \@firstoftwo
 \else \expandafter \@secondoftwo
 \fi
}%
\providecommand \natexlab [1]{#1}%
\providecommand \enquote  [1]{``#1''}%
\providecommand \bibnamefont  [1]{#1}%
\providecommand \bibfnamefont [1]{#1}%
\providecommand \citenamefont [1]{#1}%
\providecommand \href@noop [0]{\@secondoftwo}%
\providecommand \href [0]{\begingroup \@sanitize@url \@href}%
\providecommand \@href[1]{\@@startlink{#1}\@@href}%
\providecommand \@@href[1]{\endgroup#1\@@endlink}%
\providecommand \@sanitize@url [0]{\catcode `\\12\catcode `\$12\catcode `\&12\catcode `\#12\catcode `\^12\catcode `\_12\catcode `\%12\relax}%
\providecommand \@@startlink[1]{}%
\providecommand \@@endlink[0]{}%
\providecommand \url  [0]{\begingroup\@sanitize@url \@url }%
\providecommand \@url [1]{\endgroup\@href {#1}{\urlprefix }}%
\providecommand \urlprefix  [0]{URL }%
\providecommand \Eprint [0]{\href }%
\providecommand \doibase [0]{https://doi.org/}%
\providecommand \selectlanguage [0]{\@gobble}%
\providecommand \bibinfo  [0]{\@secondoftwo}%
\providecommand \bibfield  [0]{\@secondoftwo}%
\providecommand \translation [1]{[#1]}%
\providecommand \BibitemOpen [0]{}%
\providecommand \bibitemStop [0]{}%
\providecommand \bibitemNoStop [0]{.\EOS\space}%
\providecommand \EOS [0]{\spacefactor3000\relax}%
\providecommand \BibitemShut  [1]{\csname bibitem#1\endcsname}%
\let\auto@bib@innerbib\@empty
\bibitem [{\citenamefont {Nowack}\ \emph {et~al.}(2007)\citenamefont {Nowack}, \citenamefont {Koppens}, \citenamefont {Nazarov},\ and\ \citenamefont {Vandersypen}}]{Nowack}%
  \BibitemOpen
  \bibfield  {author} {\bibinfo {author} {\bibfnamefont {K.~C.}\ \bibnamefont {Nowack}}, \bibinfo {author} {\bibfnamefont {F.~H.~L.}\ \bibnamefont {Koppens}}, \bibinfo {author} {\bibfnamefont {Y.~V.}\ \bibnamefont {Nazarov}},\ and\ \bibinfo {author} {\bibfnamefont {L.~M.~K.}\ \bibnamefont {Vandersypen}},\ }\bibfield  {title} {\bibinfo {title} {Coherent control of a single electron spin with electric fields},\ }\href {https://doi.org/10.1126/science.1148092} {\bibfield  {journal} {\bibinfo  {journal} {Science}\ }\textbf {\bibinfo {volume} {318}},\ \bibinfo {pages} {1430–1433} (\bibinfo {year} {2007})}\BibitemShut {NoStop}%
\bibitem [{\citenamefont {Pioro-Ladrière}\ \emph {et~al.}(2008)\citenamefont {Pioro-Ladrière}, \citenamefont {Obata}, \citenamefont {Tokura}, \citenamefont {Shin}, \citenamefont {Kubo}, \citenamefont {Yoshida}, \citenamefont {Taniyama},\ and\ \citenamefont {Tarucha}}]{PioroLadrire}%
  \BibitemOpen
  \bibfield  {author} {\bibinfo {author} {\bibfnamefont {M.}~\bibnamefont {Pioro-Ladrière}}, \bibinfo {author} {\bibfnamefont {T.}~\bibnamefont {Obata}}, \bibinfo {author} {\bibfnamefont {Y.}~\bibnamefont {Tokura}}, \bibinfo {author} {\bibfnamefont {Y.-S.}\ \bibnamefont {Shin}}, \bibinfo {author} {\bibfnamefont {T.}~\bibnamefont {Kubo}}, \bibinfo {author} {\bibfnamefont {K.}~\bibnamefont {Yoshida}}, \bibinfo {author} {\bibfnamefont {T.}~\bibnamefont {Taniyama}},\ and\ \bibinfo {author} {\bibfnamefont {S.}~\bibnamefont {Tarucha}},\ }\bibfield  {title} {\bibinfo {title} {Electrically driven single-electron spin resonance in a slanting zeeman field},\ }\href {https://doi.org/10.1038/nphys1053} {\bibfield  {journal} {\bibinfo  {journal} {Nat. Phys.}\ }\textbf {\bibinfo {volume} {4}},\ \bibinfo {pages} {776–779} (\bibinfo {year} {2008})}\BibitemShut {NoStop}%
\bibitem [{\citenamefont {Golovach}\ \emph {et~al.}(2006)\citenamefont {Golovach}, \citenamefont {Borhani},\ and\ \citenamefont {Loss}}]{Golovach}%
  \BibitemOpen
  \bibfield  {author} {\bibinfo {author} {\bibfnamefont {V.~N.}\ \bibnamefont {Golovach}}, \bibinfo {author} {\bibfnamefont {M.}~\bibnamefont {Borhani}},\ and\ \bibinfo {author} {\bibfnamefont {D.}~\bibnamefont {Loss}},\ }\bibfield  {title} {\bibinfo {title} {Electric-dipole-induced spin resonance in quantum dots},\ }\href {https://doi.org/10.1103/physrevb.74.165319} {\bibfield  {journal} {\bibinfo  {journal} {Phys. Rev. B}\ }\textbf {\bibinfo {volume} {74}},\ \bibinfo {pages} {165319} (\bibinfo {year} {2006})}\BibitemShut {NoStop}%
\bibitem [{\citenamefont {Burkard}\ \emph {et~al.}(1999)\citenamefont {Burkard}, \citenamefont {Loss},\ and\ \citenamefont {DiVincenzo}}]{Burkard}%
  \BibitemOpen
  \bibfield  {author} {\bibinfo {author} {\bibfnamefont {G.}~\bibnamefont {Burkard}}, \bibinfo {author} {\bibfnamefont {D.}~\bibnamefont {Loss}},\ and\ \bibinfo {author} {\bibfnamefont {D.~P.}\ \bibnamefont {DiVincenzo}},\ }\bibfield  {title} {\bibinfo {title} {Coupled quantum dots as quantum gates},\ }\href {https://doi.org/10.1103/physrevb.59.2070} {\bibfield  {journal} {\bibinfo  {journal} {Phys. Rev. B}\ }\textbf {\bibinfo {volume} {59}},\ \bibinfo {pages} {2070–2078} (\bibinfo {year} {1999})}\BibitemShut {NoStop}%
\bibitem [{\citenamefont {Brunner}\ \emph {et~al.}(2011)\citenamefont {Brunner}, \citenamefont {Shin}, \citenamefont {Obata}, \citenamefont {Pioro-Ladrière}, \citenamefont {Kubo}, \citenamefont {Yoshida}, \citenamefont {Taniyama}, \citenamefont {Tokura},\ and\ \citenamefont {Tarucha}}]{Brunner}%
  \BibitemOpen
  \bibfield  {author} {\bibinfo {author} {\bibfnamefont {R.}~\bibnamefont {Brunner}}, \bibinfo {author} {\bibfnamefont {Y.-S.}\ \bibnamefont {Shin}}, \bibinfo {author} {\bibfnamefont {T.}~\bibnamefont {Obata}}, \bibinfo {author} {\bibfnamefont {M.}~\bibnamefont {Pioro-Ladrière}}, \bibinfo {author} {\bibfnamefont {T.}~\bibnamefont {Kubo}}, \bibinfo {author} {\bibfnamefont {K.}~\bibnamefont {Yoshida}}, \bibinfo {author} {\bibfnamefont {T.}~\bibnamefont {Taniyama}}, \bibinfo {author} {\bibfnamefont {Y.}~\bibnamefont {Tokura}},\ and\ \bibinfo {author} {\bibfnamefont {S.}~\bibnamefont {Tarucha}},\ }\bibfield  {title} {\bibinfo {title} {Two-qubit gate of combined single-spin rotation and interdot spin exchange in a double quantum dot},\ }\href {https://doi.org/10.1103/physrevlett.107.146801} {\bibfield  {journal} {\bibinfo  {journal} {Phys. Rev. Lett.}\ }\textbf {\bibinfo {volume} {107}},\ \bibinfo {pages} {146801} (\bibinfo {year} {2011})}\BibitemShut {NoStop}%
\bibitem [{\citenamefont {Zajac}\ \emph {et~al.}(2016{\natexlab{a}})\citenamefont {Zajac}, \citenamefont {Hazard}, \citenamefont {Mi}, \citenamefont {Nielsen},\ and\ \citenamefont {Petta}}]{zajac}%
  \BibitemOpen
  \bibfield  {author} {\bibinfo {author} {\bibfnamefont {D.~M.}\ \bibnamefont {Zajac}}, \bibinfo {author} {\bibfnamefont {T.~M.}\ \bibnamefont {Hazard}}, \bibinfo {author} {\bibfnamefont {X.}~\bibnamefont {Mi}}, \bibinfo {author} {\bibfnamefont {E.}~\bibnamefont {Nielsen}},\ and\ \bibinfo {author} {\bibfnamefont {J.~R.}\ \bibnamefont {Petta}},\ }\bibfield  {title} {\bibinfo {title} {Scalable gate architecture for a one-dimensional array of semiconductor spin qubits},\ }\href {https://doi.org/10.1103/PhysRevApplied.6.054013} {\bibfield  {journal} {\bibinfo  {journal} {Phys. Rev. Appl.}\ }\textbf {\bibinfo {volume} {6}},\ \bibinfo {pages} {054013} (\bibinfo {year} {2016}{\natexlab{a}})}\BibitemShut {NoStop}%
\bibitem [{\citenamefont {Yoneda}\ \emph {et~al.}(2018)\citenamefont {Yoneda}, \citenamefont {Takeda}, \citenamefont {Otsuka}, \citenamefont {Nakajima}, \citenamefont {Delbecq}, \citenamefont {Allison}, \citenamefont {Honda}, \citenamefont {Kodera}, \citenamefont {Oda}, \citenamefont {Hoshi} \emph {et~al.}}]{yoneda}%
  \BibitemOpen
  \bibfield  {author} {\bibinfo {author} {\bibfnamefont {J.}~\bibnamefont {Yoneda}}, \bibinfo {author} {\bibfnamefont {K.}~\bibnamefont {Takeda}}, \bibinfo {author} {\bibfnamefont {T.}~\bibnamefont {Otsuka}}, \bibinfo {author} {\bibfnamefont {T.}~\bibnamefont {Nakajima}}, \bibinfo {author} {\bibfnamefont {M.~R.}\ \bibnamefont {Delbecq}}, \bibinfo {author} {\bibfnamefont {G.}~\bibnamefont {Allison}}, \bibinfo {author} {\bibfnamefont {T.}~\bibnamefont {Honda}}, \bibinfo {author} {\bibfnamefont {T.}~\bibnamefont {Kodera}}, \bibinfo {author} {\bibfnamefont {S.}~\bibnamefont {Oda}}, \bibinfo {author} {\bibfnamefont {Y.}~\bibnamefont {Hoshi}}, \emph {et~al.},\ }\bibfield  {title} {\bibinfo {title} {A quantum-dot spin qubit with coherence limited by charge noise and fidelity higher than 99.9\%},\ }\href {https://doi.org/10.1038/s41565-017-0014-x} {\bibfield  {journal} {\bibinfo  {journal} {Nat. Nanotechnol.}\ }\textbf {\bibinfo {volume} {13}},\ \bibinfo {pages} {102} (\bibinfo {year} {2018})}\BibitemShut {NoStop}%
\bibitem [{\citenamefont {Watson}\ \emph {et~al.}(2018)\citenamefont {Watson}, \citenamefont {Philips}, \citenamefont {Kawakami}, \citenamefont {Ward}, \citenamefont {Scarlino}, \citenamefont {Veldhorst}, \citenamefont {Savage}, \citenamefont {Lagally}, \citenamefont {Friesen}, \citenamefont {Coppersmith}, \citenamefont {Eriksson},\ and\ \citenamefont {Vandersypen}}]{Watson}%
  \BibitemOpen
  \bibfield  {author} {\bibinfo {author} {\bibfnamefont {T.~F.}\ \bibnamefont {Watson}}, \bibinfo {author} {\bibfnamefont {S.~G.~J.}\ \bibnamefont {Philips}}, \bibinfo {author} {\bibfnamefont {E.}~\bibnamefont {Kawakami}}, \bibinfo {author} {\bibfnamefont {D.~R.}\ \bibnamefont {Ward}}, \bibinfo {author} {\bibfnamefont {P.}~\bibnamefont {Scarlino}}, \bibinfo {author} {\bibfnamefont {M.}~\bibnamefont {Veldhorst}}, \bibinfo {author} {\bibfnamefont {D.~E.}\ \bibnamefont {Savage}}, \bibinfo {author} {\bibfnamefont {M.~G.}\ \bibnamefont {Lagally}}, \bibinfo {author} {\bibfnamefont {M.}~\bibnamefont {Friesen}}, \bibinfo {author} {\bibfnamefont {S.~N.}\ \bibnamefont {Coppersmith}}, \bibinfo {author} {\bibfnamefont {M.~A.}\ \bibnamefont {Eriksson}},\ and\ \bibinfo {author} {\bibfnamefont {L.~M.~K.}\ \bibnamefont {Vandersypen}},\ }\bibfield  {title} {\bibinfo {title} {A programmable two-qubit quantum processor in silicon},\ }\href {https://doi.org/10.1038/nature25766} {\bibfield  {journal} {\bibinfo  {journal}
  {Nature}\ }\textbf {\bibinfo {volume} {555}},\ \bibinfo {pages} {633–637} (\bibinfo {year} {2018})}\BibitemShut {NoStop}%
\bibitem [{\citenamefont {Khaetskii}\ \emph {et~al.}(2002)\citenamefont {Khaetskii}, \citenamefont {Loss},\ and\ \citenamefont {Glazman}}]{Khaetskii}%
  \BibitemOpen
  \bibfield  {author} {\bibinfo {author} {\bibfnamefont {A.~V.}\ \bibnamefont {Khaetskii}}, \bibinfo {author} {\bibfnamefont {D.}~\bibnamefont {Loss}},\ and\ \bibinfo {author} {\bibfnamefont {L.}~\bibnamefont {Glazman}},\ }\bibfield  {title} {\bibinfo {title} {Electron spin decoherence in quantum dots due to interaction with nuclei},\ }\href {https://doi.org/10.1103/physrevlett.88.186802} {\bibfield  {journal} {\bibinfo  {journal} {Phys. Rev. Lett.}\ }\textbf {\bibinfo {volume} {88}},\ \bibinfo {pages} {186802} (\bibinfo {year} {2002})}\BibitemShut {NoStop}%
\bibitem [{\citenamefont {Bluhm}\ \emph {et~al.}(2010)\citenamefont {Bluhm}, \citenamefont {Foletti}, \citenamefont {Neder}, \citenamefont {Rudner}, \citenamefont {Mahalu}, \citenamefont {Umansky},\ and\ \citenamefont {Yacoby}}]{Bluhm}%
  \BibitemOpen
  \bibfield  {author} {\bibinfo {author} {\bibfnamefont {H.}~\bibnamefont {Bluhm}}, \bibinfo {author} {\bibfnamefont {S.}~\bibnamefont {Foletti}}, \bibinfo {author} {\bibfnamefont {I.}~\bibnamefont {Neder}}, \bibinfo {author} {\bibfnamefont {M.}~\bibnamefont {Rudner}}, \bibinfo {author} {\bibfnamefont {D.}~\bibnamefont {Mahalu}}, \bibinfo {author} {\bibfnamefont {V.}~\bibnamefont {Umansky}},\ and\ \bibinfo {author} {\bibfnamefont {A.}~\bibnamefont {Yacoby}},\ }\bibfield  {title} {\bibinfo {title} {Dephasing time of gaas electron-spin qubits coupled to a nuclear bath exceeding 200~$\mu$s},\ }\href {https://doi.org/10.1038/nphys1856} {\bibfield  {journal} {\bibinfo  {journal} {Nat. Phys.}\ }\textbf {\bibinfo {volume} {7}},\ \bibinfo {pages} {109} (\bibinfo {year} {2010})}\BibitemShut {NoStop}%
\bibitem [{\citenamefont {Itoh}\ \emph {et~al.}(1993)\citenamefont {Itoh}, \citenamefont {Hansen}, \citenamefont {Haller}, \citenamefont {Farmer}, \citenamefont {Ozhogin}, \citenamefont {Rudnev},\ and\ \citenamefont {Tikhomirov}}]{Itoh}%
  \BibitemOpen
  \bibfield  {author} {\bibinfo {author} {\bibfnamefont {K.}~\bibnamefont {Itoh}}, \bibinfo {author} {\bibfnamefont {W.}~\bibnamefont {Hansen}}, \bibinfo {author} {\bibfnamefont {E.}~\bibnamefont {Haller}}, \bibinfo {author} {\bibfnamefont {J.}~\bibnamefont {Farmer}}, \bibinfo {author} {\bibfnamefont {V.}~\bibnamefont {Ozhogin}}, \bibinfo {author} {\bibfnamefont {A.}~\bibnamefont {Rudnev}},\ and\ \bibinfo {author} {\bibfnamefont {A.}~\bibnamefont {Tikhomirov}},\ }\bibfield  {title} {\bibinfo {title} {High purity isotopically enriched 70ge and 74ge single crystals: Isotope separation, growth, and properties},\ }\href {https://doi.org/10.1557/jmr.1993.1341} {\bibfield  {journal} {\bibinfo  {journal} {J. Mater. Res.}\ }\textbf {\bibinfo {volume} {8}},\ \bibinfo {pages} {1341–1347} (\bibinfo {year} {1993})}\BibitemShut {NoStop}%
\bibitem [{\citenamefont {Lodari}\ \emph {et~al.}(2019)\citenamefont {Lodari}, \citenamefont {Tosato}, \citenamefont {Sabbagh}, \citenamefont {Schubert}, \citenamefont {Capellini}, \citenamefont {Sammak}, \citenamefont {Veldhorst},\ and\ \citenamefont {Scappucci}}]{Lodari}%
  \BibitemOpen
  \bibfield  {author} {\bibinfo {author} {\bibfnamefont {M.}~\bibnamefont {Lodari}}, \bibinfo {author} {\bibfnamefont {A.}~\bibnamefont {Tosato}}, \bibinfo {author} {\bibfnamefont {D.}~\bibnamefont {Sabbagh}}, \bibinfo {author} {\bibfnamefont {M.~A.}\ \bibnamefont {Schubert}}, \bibinfo {author} {\bibfnamefont {G.}~\bibnamefont {Capellini}}, \bibinfo {author} {\bibfnamefont {A.}~\bibnamefont {Sammak}}, \bibinfo {author} {\bibfnamefont {M.}~\bibnamefont {Veldhorst}},\ and\ \bibinfo {author} {\bibfnamefont {G.}~\bibnamefont {Scappucci}},\ }\bibfield  {title} {\bibinfo {title} {Light effective hole mass in undoped ge/sige quantum wells},\ }\href {https://doi.org/10.1103/physrevb.100.041304} {\bibfield  {journal} {\bibinfo  {journal} {Phys. Rev. B}\ }\textbf {\bibinfo {volume} {100}},\ \bibinfo {pages} {100} (\bibinfo {year} {2019})}\BibitemShut {NoStop}%
\bibitem [{\citenamefont {Scappucci}\ \emph {et~al.}(2020)\citenamefont {Scappucci}, \citenamefont {Kloeffel}, \citenamefont {Zwanenburg}, \citenamefont {Loss}, \citenamefont {Myronov}, \citenamefont {Zhang}, \citenamefont {De~Franceschi}, \citenamefont {Katsaros},\ and\ \citenamefont {Veldhorst}}]{Scappucci}%
  \BibitemOpen
  \bibfield  {author} {\bibinfo {author} {\bibfnamefont {G.}~\bibnamefont {Scappucci}}, \bibinfo {author} {\bibfnamefont {C.}~\bibnamefont {Kloeffel}}, \bibinfo {author} {\bibfnamefont {F.~A.}\ \bibnamefont {Zwanenburg}}, \bibinfo {author} {\bibfnamefont {D.}~\bibnamefont {Loss}}, \bibinfo {author} {\bibfnamefont {M.}~\bibnamefont {Myronov}}, \bibinfo {author} {\bibfnamefont {J.-J.}\ \bibnamefont {Zhang}}, \bibinfo {author} {\bibfnamefont {S.}~\bibnamefont {De~Franceschi}}, \bibinfo {author} {\bibfnamefont {G.}~\bibnamefont {Katsaros}},\ and\ \bibinfo {author} {\bibfnamefont {M.}~\bibnamefont {Veldhorst}},\ }\bibfield  {title} {\bibinfo {title} {The germanium quantum information route},\ }\href {https://doi.org/10.1038/s41578-020-00262-z} {\bibfield  {journal} {\bibinfo  {journal} {Nat. Rev. Mater.}\ }\textbf {\bibinfo {volume} {6}},\ \bibinfo {pages} {926–943} (\bibinfo {year} {2020})}\BibitemShut {NoStop}%
\bibitem [{\citenamefont {Hendrickx}\ \emph {et~al.}(2020)\citenamefont {Hendrickx}, \citenamefont {Franke}, \citenamefont {Sammak}, \citenamefont {Scappucci},\ and\ \citenamefont {Veldhorst}}]{Hendrickx}%
  \BibitemOpen
  \bibfield  {author} {\bibinfo {author} {\bibfnamefont {N.~W.}\ \bibnamefont {Hendrickx}}, \bibinfo {author} {\bibfnamefont {D.~P.}\ \bibnamefont {Franke}}, \bibinfo {author} {\bibfnamefont {A.}~\bibnamefont {Sammak}}, \bibinfo {author} {\bibfnamefont {G.}~\bibnamefont {Scappucci}},\ and\ \bibinfo {author} {\bibfnamefont {M.}~\bibnamefont {Veldhorst}},\ }\bibfield  {title} {\bibinfo {title} {Fast two-qubit logic with holes in germanium},\ }\href {https://doi.org/10.1038/s41586-019-1919-3} {\bibfield  {journal} {\bibinfo  {journal} {Nature}\ }\textbf {\bibinfo {volume} {577}},\ \bibinfo {pages} {487–491} (\bibinfo {year} {2020})}\BibitemShut {NoStop}%
\bibitem [{\citenamefont {Lawrie}\ \emph {et~al.}(2023)\citenamefont {Lawrie}, \citenamefont {Rimbach-Russ}, \citenamefont {Riggelen}, \citenamefont {Hendrickx}, \citenamefont {Snoo}, \citenamefont {Sammak}, \citenamefont {Scappucci}, \citenamefont {Helsen},\ and\ \citenamefont {Veldhorst}}]{Lawrie}%
  \BibitemOpen
  \bibfield  {author} {\bibinfo {author} {\bibfnamefont {W.~I.~L.}\ \bibnamefont {Lawrie}}, \bibinfo {author} {\bibfnamefont {M.}~\bibnamefont {Rimbach-Russ}}, \bibinfo {author} {\bibfnamefont {F.~v.}\ \bibnamefont {Riggelen}}, \bibinfo {author} {\bibfnamefont {N.~W.}\ \bibnamefont {Hendrickx}}, \bibinfo {author} {\bibfnamefont {S.~L.~d.}\ \bibnamefont {Snoo}}, \bibinfo {author} {\bibfnamefont {A.}~\bibnamefont {Sammak}}, \bibinfo {author} {\bibfnamefont {G.}~\bibnamefont {Scappucci}}, \bibinfo {author} {\bibfnamefont {J.}~\bibnamefont {Helsen}},\ and\ \bibinfo {author} {\bibfnamefont {M.}~\bibnamefont {Veldhorst}},\ }\bibfield  {title} {\bibinfo {title} {Simultaneous single-qubit driving of semiconductor spin qubits at the fault-tolerant threshold},\ }\href {https://doi.org/10.1038/s41467-023-39334-3} {\bibfield  {journal} {\bibinfo  {journal} {Nat. Commun.}\ }\textbf {\bibinfo {volume} {14}},\ \bibinfo {pages} {3617} (\bibinfo {year} {2023})}\BibitemShut {NoStop}%
\bibitem [{\citenamefont {Jirovec}\ \emph {et~al.}(2021)\citenamefont {Jirovec}, \citenamefont {Hofmann}, \citenamefont {Ballabio}, \citenamefont {Mutter}, \citenamefont {Tavani}, \citenamefont {Botifoll}, \citenamefont {Crippa}, \citenamefont {Kukucka}, \citenamefont {Sagi}, \citenamefont {Martins}, \citenamefont {Saez-Mollejo}, \citenamefont {Prieto}, \citenamefont {Borovkov}, \citenamefont {Arbiol}, \citenamefont {Chrastina}, \citenamefont {Isella},\ and\ \citenamefont {Katsaros}}]{Jirovec}%
  \BibitemOpen
  \bibfield  {author} {\bibinfo {author} {\bibfnamefont {D.}~\bibnamefont {Jirovec}}, \bibinfo {author} {\bibfnamefont {A.}~\bibnamefont {Hofmann}}, \bibinfo {author} {\bibfnamefont {A.}~\bibnamefont {Ballabio}}, \bibinfo {author} {\bibfnamefont {P.~M.}\ \bibnamefont {Mutter}}, \bibinfo {author} {\bibfnamefont {G.}~\bibnamefont {Tavani}}, \bibinfo {author} {\bibfnamefont {M.}~\bibnamefont {Botifoll}}, \bibinfo {author} {\bibfnamefont {A.}~\bibnamefont {Crippa}}, \bibinfo {author} {\bibfnamefont {J.}~\bibnamefont {Kukucka}}, \bibinfo {author} {\bibfnamefont {O.}~\bibnamefont {Sagi}}, \bibinfo {author} {\bibfnamefont {F.}~\bibnamefont {Martins}}, \bibinfo {author} {\bibfnamefont {J.}~\bibnamefont {Saez-Mollejo}}, \bibinfo {author} {\bibfnamefont {I.}~\bibnamefont {Prieto}}, \bibinfo {author} {\bibfnamefont {M.}~\bibnamefont {Borovkov}}, \bibinfo {author} {\bibfnamefont {J.}~\bibnamefont {Arbiol}}, \bibinfo {author} {\bibfnamefont {D.}~\bibnamefont {Chrastina}}, \bibinfo {author} {\bibfnamefont {G.}~\bibnamefont
  {Isella}},\ and\ \bibinfo {author} {\bibfnamefont {G.}~\bibnamefont {Katsaros}},\ }\bibfield  {title} {\bibinfo {title} {A singlet-triplet hole spin qubit in planar ge},\ }\href {https://doi.org/10.1038/s41563-021-01022-2} {\bibfield  {journal} {\bibinfo  {journal} {Nat. Mater.}\ }\textbf {\bibinfo {volume} {20}},\ \bibinfo {pages} {1106–1112} (\bibinfo {year} {2021})}\BibitemShut {NoStop}%
\bibitem [{\citenamefont {Hendrickx}\ \emph {et~al.}(2021)\citenamefont {Hendrickx}, \citenamefont {Lawrie}, \citenamefont {Russ}, \citenamefont {van Riggelen}, \citenamefont {de~Snoo}, \citenamefont {Schouten}, \citenamefont {Sammak}, \citenamefont {Scappucci},\ and\ \citenamefont {Veldhorst}}]{Hendrickx2}%
  \BibitemOpen
  \bibfield  {author} {\bibinfo {author} {\bibfnamefont {N.~W.}\ \bibnamefont {Hendrickx}}, \bibinfo {author} {\bibfnamefont {W.~I.~L.}\ \bibnamefont {Lawrie}}, \bibinfo {author} {\bibfnamefont {M.}~\bibnamefont {Russ}}, \bibinfo {author} {\bibfnamefont {F.}~\bibnamefont {van Riggelen}}, \bibinfo {author} {\bibfnamefont {S.~L.}\ \bibnamefont {de~Snoo}}, \bibinfo {author} {\bibfnamefont {R.~N.}\ \bibnamefont {Schouten}}, \bibinfo {author} {\bibfnamefont {A.}~\bibnamefont {Sammak}}, \bibinfo {author} {\bibfnamefont {G.}~\bibnamefont {Scappucci}},\ and\ \bibinfo {author} {\bibfnamefont {M.}~\bibnamefont {Veldhorst}},\ }\bibfield  {title} {\bibinfo {title} {A four-qubit germanium quantum processor},\ }\href {https://doi.org/10.1038/s41586-021-03332-6} {\bibfield  {journal} {\bibinfo  {journal} {Nature}\ }\textbf {\bibinfo {volume} {591}},\ \bibinfo {pages} {580–585} (\bibinfo {year} {2021})}\BibitemShut {NoStop}%
\bibitem [{\citenamefont {van Riggelen}\ \emph {et~al.}(2022)\citenamefont {van Riggelen}, \citenamefont {Lawrie}, \citenamefont {Russ}, \citenamefont {Hendrickx}, \citenamefont {Sammak}, \citenamefont {Rispler}, \citenamefont {Terhal}, \citenamefont {Scappucci},\ and\ \citenamefont {Veldhorst}}]{vanRiggelen}%
  \BibitemOpen
  \bibfield  {author} {\bibinfo {author} {\bibfnamefont {F.}~\bibnamefont {van Riggelen}}, \bibinfo {author} {\bibfnamefont {W.~I.~L.}\ \bibnamefont {Lawrie}}, \bibinfo {author} {\bibfnamefont {M.}~\bibnamefont {Russ}}, \bibinfo {author} {\bibfnamefont {N.~W.}\ \bibnamefont {Hendrickx}}, \bibinfo {author} {\bibfnamefont {A.}~\bibnamefont {Sammak}}, \bibinfo {author} {\bibfnamefont {M.}~\bibnamefont {Rispler}}, \bibinfo {author} {\bibfnamefont {B.~M.}\ \bibnamefont {Terhal}}, \bibinfo {author} {\bibfnamefont {G.}~\bibnamefont {Scappucci}},\ and\ \bibinfo {author} {\bibfnamefont {M.}~\bibnamefont {Veldhorst}},\ }\bibfield  {title} {\bibinfo {title} {Phase flip code with semiconductor spin qubits},\ }\href {https://doi.org/10.1038/s41534-022-00639-8} {\bibfield  {journal} {\bibinfo  {journal} {npj Quantum Inf.}\ }\textbf {\bibinfo {volume} {8}},\ \bibinfo {pages} {124} (\bibinfo {year} {2022})}\BibitemShut {NoStop}%
\bibitem [{\citenamefont {Borsoi}\ \emph {et~al.}(2023)\citenamefont {Borsoi}, \citenamefont {Hendrickx}, \citenamefont {John}, \citenamefont {Meyer}, \citenamefont {Motz}, \citenamefont {van Riggelen}, \citenamefont {Sammak}, \citenamefont {de~Snoo}, \citenamefont {Scappucci},\ and\ \citenamefont {Veldhorst}}]{Borsoi}%
  \BibitemOpen
  \bibfield  {author} {\bibinfo {author} {\bibfnamefont {F.}~\bibnamefont {Borsoi}}, \bibinfo {author} {\bibfnamefont {N.~W.}\ \bibnamefont {Hendrickx}}, \bibinfo {author} {\bibfnamefont {V.}~\bibnamefont {John}}, \bibinfo {author} {\bibfnamefont {M.}~\bibnamefont {Meyer}}, \bibinfo {author} {\bibfnamefont {S.}~\bibnamefont {Motz}}, \bibinfo {author} {\bibfnamefont {F.}~\bibnamefont {van Riggelen}}, \bibinfo {author} {\bibfnamefont {A.}~\bibnamefont {Sammak}}, \bibinfo {author} {\bibfnamefont {S.~L.}\ \bibnamefont {de~Snoo}}, \bibinfo {author} {\bibfnamefont {G.}~\bibnamefont {Scappucci}},\ and\ \bibinfo {author} {\bibfnamefont {M.}~\bibnamefont {Veldhorst}},\ }\bibfield  {title} {\bibinfo {title} {Shared control of a 16 semiconductor quantum dot crossbar array},\ }\href {https://doi.org/10.1038/s41565-023-01491-3} {\bibfield  {journal} {\bibinfo  {journal} {Nat. Nanotechnol.}\ }\textbf {\bibinfo {volume} {19}},\ \bibinfo {pages} {21–27} (\bibinfo {year} {2023})}\BibitemShut {NoStop}%
\bibitem [{\citenamefont {Zhang}\ \emph {et~al.}(2021)\citenamefont {Zhang}, \citenamefont {Liu}, \citenamefont {Gao}, \citenamefont {Xu}, \citenamefont {Wang}, \citenamefont {Zhang}, \citenamefont {Cao}, \citenamefont {Wang}, \citenamefont {Zhang}, \citenamefont {Hu}, \citenamefont {Li},\ and\ \citenamefont {Guo}}]{Zhang}%
  \BibitemOpen
  \bibfield  {author} {\bibinfo {author} {\bibfnamefont {T.}~\bibnamefont {Zhang}}, \bibinfo {author} {\bibfnamefont {H.}~\bibnamefont {Liu}}, \bibinfo {author} {\bibfnamefont {F.}~\bibnamefont {Gao}}, \bibinfo {author} {\bibfnamefont {G.}~\bibnamefont {Xu}}, \bibinfo {author} {\bibfnamefont {K.}~\bibnamefont {Wang}}, \bibinfo {author} {\bibfnamefont {X.}~\bibnamefont {Zhang}}, \bibinfo {author} {\bibfnamefont {G.}~\bibnamefont {Cao}}, \bibinfo {author} {\bibfnamefont {T.}~\bibnamefont {Wang}}, \bibinfo {author} {\bibfnamefont {J.}~\bibnamefont {Zhang}}, \bibinfo {author} {\bibfnamefont {X.}~\bibnamefont {Hu}}, \bibinfo {author} {\bibfnamefont {H.-O.}\ \bibnamefont {Li}},\ and\ \bibinfo {author} {\bibfnamefont {G.-P.}\ \bibnamefont {Guo}},\ }\bibfield  {title} {\bibinfo {title} {Anisotropic g-factor and spin–orbit field in a germanium hut wire double quantum dot},\ }\href {https://doi.org/10.1021/acs.nanolett.1c00263} {\bibfield  {journal} {\bibinfo  {journal} {Nano Lett.}\ }\textbf {\bibinfo {volume}
  {21}},\ \bibinfo {pages} {3835–3842} (\bibinfo {year} {2021})}\BibitemShut {NoStop}%
\bibitem [{\citenamefont {Jirovec}\ \emph {et~al.}(2022)\citenamefont {Jirovec}, \citenamefont {Mutter}, \citenamefont {Hofmann}, \citenamefont {Crippa}, \citenamefont {Rychetsky}, \citenamefont {Craig}, \citenamefont {Kukucka}, \citenamefont {Martins}, \citenamefont {Ballabio}, \citenamefont {Ares}, \citenamefont {Chrastina}, \citenamefont {Isella}, \citenamefont {Burkard},\ and\ \citenamefont {Katsaros}}]{Jirovec2}%
  \BibitemOpen
  \bibfield  {author} {\bibinfo {author} {\bibfnamefont {D.}~\bibnamefont {Jirovec}}, \bibinfo {author} {\bibfnamefont {P.~M.}\ \bibnamefont {Mutter}}, \bibinfo {author} {\bibfnamefont {A.}~\bibnamefont {Hofmann}}, \bibinfo {author} {\bibfnamefont {A.}~\bibnamefont {Crippa}}, \bibinfo {author} {\bibfnamefont {M.}~\bibnamefont {Rychetsky}}, \bibinfo {author} {\bibfnamefont {D.~L.}\ \bibnamefont {Craig}}, \bibinfo {author} {\bibfnamefont {J.}~\bibnamefont {Kukucka}}, \bibinfo {author} {\bibfnamefont {F.}~\bibnamefont {Martins}}, \bibinfo {author} {\bibfnamefont {A.}~\bibnamefont {Ballabio}}, \bibinfo {author} {\bibfnamefont {N.}~\bibnamefont {Ares}}, \bibinfo {author} {\bibfnamefont {D.}~\bibnamefont {Chrastina}}, \bibinfo {author} {\bibfnamefont {G.}~\bibnamefont {Isella}}, \bibinfo {author} {\bibfnamefont {G.}~\bibnamefont {Burkard}},\ and\ \bibinfo {author} {\bibfnamefont {G.}~\bibnamefont {Katsaros}},\ }\bibfield  {title} {\bibinfo {title} {Dynamics of hole singlet-triplet qubits with large g-factor
  differences},\ }\href {https://doi.org/10.1103/physrevlett.128.126803} {\bibfield  {journal} {\bibinfo  {journal} {Phys. Rev. Lett.}\ }\textbf {\bibinfo {volume} {128}},\ \bibinfo {pages} {126803} (\bibinfo {year} {2022})}\BibitemShut {NoStop}%
\bibitem [{\citenamefont {Piot}\ \emph {et~al.}(2022)\citenamefont {Piot}, \citenamefont {Brun}, \citenamefont {Schmitt}, \citenamefont {Zihlmann}, \citenamefont {Michal}, \citenamefont {Apra}, \citenamefont {Abadillo-Uriel}, \citenamefont {Jehl}, \citenamefont {Bertrand}, \citenamefont {Niebojewski}, \citenamefont {Hutin}, \citenamefont {Vinet}, \citenamefont {Urdampilleta}, \citenamefont {Meunier}, \citenamefont {Niquet}, \citenamefont {Maurand},\ and\ \citenamefont {Franceschi}}]{Piot}%
  \BibitemOpen
  \bibfield  {author} {\bibinfo {author} {\bibfnamefont {N.}~\bibnamefont {Piot}}, \bibinfo {author} {\bibfnamefont {B.}~\bibnamefont {Brun}}, \bibinfo {author} {\bibfnamefont {V.}~\bibnamefont {Schmitt}}, \bibinfo {author} {\bibfnamefont {S.}~\bibnamefont {Zihlmann}}, \bibinfo {author} {\bibfnamefont {V.~P.}\ \bibnamefont {Michal}}, \bibinfo {author} {\bibfnamefont {A.}~\bibnamefont {Apra}}, \bibinfo {author} {\bibfnamefont {J.~C.}\ \bibnamefont {Abadillo-Uriel}}, \bibinfo {author} {\bibfnamefont {X.}~\bibnamefont {Jehl}}, \bibinfo {author} {\bibfnamefont {B.}~\bibnamefont {Bertrand}}, \bibinfo {author} {\bibfnamefont {H.}~\bibnamefont {Niebojewski}}, \bibinfo {author} {\bibfnamefont {L.}~\bibnamefont {Hutin}}, \bibinfo {author} {\bibfnamefont {M.}~\bibnamefont {Vinet}}, \bibinfo {author} {\bibfnamefont {M.}~\bibnamefont {Urdampilleta}}, \bibinfo {author} {\bibfnamefont {T.}~\bibnamefont {Meunier}}, \bibinfo {author} {\bibfnamefont {Y.-M.}\ \bibnamefont {Niquet}}, \bibinfo {author} {\bibfnamefont
  {R.}~\bibnamefont {Maurand}},\ and\ \bibinfo {author} {\bibfnamefont {S.~D.}\ \bibnamefont {Franceschi}},\ }\bibfield  {title} {\bibinfo {title} {A single hole spin with enhanced coherence in natural silicon},\ }\href {https://doi.org/10.1038/s41565-022-01196-z} {\bibfield  {journal} {\bibinfo  {journal} {Nat. Nanotechnol.}\ }\textbf {\bibinfo {volume} {17}},\ \bibinfo {pages} {1072–1077} (\bibinfo {year} {2022})}\BibitemShut {NoStop}%
\bibitem [{\citenamefont {Scher\"{u}bl}\ \emph {et~al.}(2019)\citenamefont {Scher\"{u}bl}, \citenamefont {Pályi}, \citenamefont {Frank}, \citenamefont {Lukács}, \citenamefont {F\"{u}l\"{o}p}, \citenamefont {F\"{u}l\"{o}p}, \citenamefont {Nygård}, \citenamefont {Watanabe}, \citenamefont {Taniguchi}, \citenamefont {Zaránd},\ and\ \citenamefont {Csonka}}]{Scherbl}%
  \BibitemOpen
  \bibfield  {author} {\bibinfo {author} {\bibfnamefont {Z.}~\bibnamefont {Scher\"{u}bl}}, \bibinfo {author} {\bibfnamefont {A.}~\bibnamefont {Pályi}}, \bibinfo {author} {\bibfnamefont {G.}~\bibnamefont {Frank}}, \bibinfo {author} {\bibfnamefont {I.~E.}\ \bibnamefont {Lukács}}, \bibinfo {author} {\bibfnamefont {G.}~\bibnamefont {F\"{u}l\"{o}p}}, \bibinfo {author} {\bibfnamefont {B.}~\bibnamefont {F\"{u}l\"{o}p}}, \bibinfo {author} {\bibfnamefont {J.}~\bibnamefont {Nygård}}, \bibinfo {author} {\bibfnamefont {K.}~\bibnamefont {Watanabe}}, \bibinfo {author} {\bibfnamefont {T.}~\bibnamefont {Taniguchi}}, \bibinfo {author} {\bibfnamefont {G.}~\bibnamefont {Zaránd}},\ and\ \bibinfo {author} {\bibfnamefont {S.}~\bibnamefont {Csonka}},\ }\bibfield  {title} {\bibinfo {title} {Observation of spin–orbit coupling induced weyl points in a two-electron double quantum dot},\ }\href {https://doi.org/10.1038/s42005-019-0200-2} {\bibfield  {journal} {\bibinfo  {journal} {Commun. Phys.}\ }\textbf {\bibinfo {volume} {2}},\
  \bibinfo {pages} {108} (\bibinfo {year} {2019})}\BibitemShut {NoStop}%
\bibitem [{\citenamefont {Crippa}\ \emph {et~al.}(2018)\citenamefont {Crippa}, \citenamefont {Maurand}, \citenamefont {Bourdet}, \citenamefont {Kotekar-Patil}, \citenamefont {Amisse}, \citenamefont {Jehl}, \citenamefont {Sanquer}, \citenamefont {Laviéville}, \citenamefont {Bohuslavskyi}, \citenamefont {Hutin}, \citenamefont {Barraud}, \citenamefont {Vinet}, \citenamefont {Niquet},\ and\ \citenamefont {De~Franceschi}}]{Crippa}%
  \BibitemOpen
  \bibfield  {author} {\bibinfo {author} {\bibfnamefont {A.}~\bibnamefont {Crippa}}, \bibinfo {author} {\bibfnamefont {R.}~\bibnamefont {Maurand}}, \bibinfo {author} {\bibfnamefont {L.}~\bibnamefont {Bourdet}}, \bibinfo {author} {\bibfnamefont {D.}~\bibnamefont {Kotekar-Patil}}, \bibinfo {author} {\bibfnamefont {A.}~\bibnamefont {Amisse}}, \bibinfo {author} {\bibfnamefont {X.}~\bibnamefont {Jehl}}, \bibinfo {author} {\bibfnamefont {M.}~\bibnamefont {Sanquer}}, \bibinfo {author} {\bibfnamefont {R.}~\bibnamefont {Laviéville}}, \bibinfo {author} {\bibfnamefont {H.}~\bibnamefont {Bohuslavskyi}}, \bibinfo {author} {\bibfnamefont {L.}~\bibnamefont {Hutin}}, \bibinfo {author} {\bibfnamefont {S.}~\bibnamefont {Barraud}}, \bibinfo {author} {\bibfnamefont {M.}~\bibnamefont {Vinet}}, \bibinfo {author} {\bibfnamefont {Y.-M.}\ \bibnamefont {Niquet}},\ and\ \bibinfo {author} {\bibfnamefont {S.}~\bibnamefont {De~Franceschi}},\ }\bibfield  {title} {\bibinfo {title} {Electrical spin driving by g-matrix modulation in
  spin-orbit qubits},\ }\href {https://doi.org/10.1103/physrevlett.120.137702} {\bibfield  {journal} {\bibinfo  {journal} {Phys. Rev. Lett.}\ }\textbf {\bibinfo {volume} {120}},\ \bibinfo {pages} {137702} (\bibinfo {year} {2018})}\BibitemShut {NoStop}%
\bibitem [{\citenamefont {Kloeffel}\ \emph {et~al.}(2011)\citenamefont {Kloeffel}, \citenamefont {Trif},\ and\ \citenamefont {Loss}}]{Kloeffel}%
  \BibitemOpen
  \bibfield  {author} {\bibinfo {author} {\bibfnamefont {C.}~\bibnamefont {Kloeffel}}, \bibinfo {author} {\bibfnamefont {M.}~\bibnamefont {Trif}},\ and\ \bibinfo {author} {\bibfnamefont {D.}~\bibnamefont {Loss}},\ }\bibfield  {title} {\bibinfo {title} {Strong spin-orbit interaction and helical hole states in ge/si nanowires},\ }\href {https://doi.org/10.1103/physrevb.84.195314} {\bibfield  {journal} {\bibinfo  {journal} {Phys. Rev. B}\ }\textbf {\bibinfo {volume} {84}},\ \bibinfo {pages} {195314} (\bibinfo {year} {2011})}\BibitemShut {NoStop}%
\bibitem [{\citenamefont {Kloeffel}\ and\ \citenamefont {Loss}(2013)}]{Kloeffel2}%
  \BibitemOpen
  \bibfield  {author} {\bibinfo {author} {\bibfnamefont {C.}~\bibnamefont {Kloeffel}}\ and\ \bibinfo {author} {\bibfnamefont {D.}~\bibnamefont {Loss}},\ }\bibfield  {title} {\bibinfo {title} {Prospects for spin-based quantum computing in quantum dots},\ }\href {https://doi.org/10.1146/annurev-conmatphys-030212-184248} {\bibfield  {journal} {\bibinfo  {journal} {Annu. Rev. Condens. Matter Phys.}\ }\textbf {\bibinfo {volume} {4}},\ \bibinfo {pages} {51–81} (\bibinfo {year} {2013})}\BibitemShut {NoStop}%
\bibitem [{\citenamefont {Nestoklon}\ \emph {et~al.}(2008)\citenamefont {Nestoklon}, \citenamefont {Ivchenko}, \citenamefont {Jancu},\ and\ \citenamefont {Voisin}}]{Nestoklon}%
  \BibitemOpen
  \bibfield  {author} {\bibinfo {author} {\bibfnamefont {M.~O.}\ \bibnamefont {Nestoklon}}, \bibinfo {author} {\bibfnamefont {E.~L.}\ \bibnamefont {Ivchenko}}, \bibinfo {author} {\bibfnamefont {J.-M.}\ \bibnamefont {Jancu}},\ and\ \bibinfo {author} {\bibfnamefont {P.}~\bibnamefont {Voisin}},\ }\bibfield  {title} {\bibinfo {title} {Electric field effect on electron spin splitting in sige/si quantum wells},\ }\href {https://doi.org/10.1103/physrevb.77.155328} {\bibfield  {journal} {\bibinfo  {journal} {Phys. Rev. B}\ }\textbf {\bibinfo {volume} {77}},\ \bibinfo {pages} {155328} (\bibinfo {year} {2008})}\BibitemShut {NoStop}%
\bibitem [{\citenamefont {Hendrickx}\ \emph {et~al.}(2024)\citenamefont {Hendrickx}, \citenamefont {Massai}, \citenamefont {Mergenthaler}, \citenamefont {Schupp}, \citenamefont {Paredes}, \citenamefont {Bedell}, \citenamefont {Salis},\ and\ \citenamefont {Fuhrer}}]{Hendrickx3}%
  \BibitemOpen
  \bibfield  {author} {\bibinfo {author} {\bibfnamefont {N.~W.}\ \bibnamefont {Hendrickx}}, \bibinfo {author} {\bibfnamefont {L.}~\bibnamefont {Massai}}, \bibinfo {author} {\bibfnamefont {M.}~\bibnamefont {Mergenthaler}}, \bibinfo {author} {\bibfnamefont {F.~J.}\ \bibnamefont {Schupp}}, \bibinfo {author} {\bibfnamefont {S.}~\bibnamefont {Paredes}}, \bibinfo {author} {\bibfnamefont {S.~W.}\ \bibnamefont {Bedell}}, \bibinfo {author} {\bibfnamefont {G.}~\bibnamefont {Salis}},\ and\ \bibinfo {author} {\bibfnamefont {A.}~\bibnamefont {Fuhrer}},\ }\bibfield  {title} {\bibinfo {title} {Sweet-spot operation of a germanium hole spin qubit with highly anisotropic noise sensitivity},\ }\href {https://doi.org/10.1038/s41563-024-01857-5} {\bibfield  {journal} {\bibinfo  {journal} {Nat. Mater.}\ }\textbf {\bibinfo {volume} {23}},\ \bibinfo {pages} {920–927} (\bibinfo {year} {2024})}\BibitemShut {NoStop}%
\bibitem [{\citenamefont {Wang}\ \emph {et~al.}(2021)\citenamefont {Wang}, \citenamefont {Marcellina}, \citenamefont {Hamilton}, \citenamefont {Cullen}, \citenamefont {Rogge}, \citenamefont {Salfi},\ and\ \citenamefont {Culcer}}]{Wang2021}%
  \BibitemOpen
  \bibfield  {author} {\bibinfo {author} {\bibfnamefont {Z.}~\bibnamefont {Wang}}, \bibinfo {author} {\bibfnamefont {E.}~\bibnamefont {Marcellina}}, \bibinfo {author} {\bibfnamefont {A.~R.}\ \bibnamefont {Hamilton}}, \bibinfo {author} {\bibfnamefont {J.~H.}\ \bibnamefont {Cullen}}, \bibinfo {author} {\bibfnamefont {S.}~\bibnamefont {Rogge}}, \bibinfo {author} {\bibfnamefont {J.}~\bibnamefont {Salfi}},\ and\ \bibinfo {author} {\bibfnamefont {D.}~\bibnamefont {Culcer}},\ }\bibfield  {title} {\bibinfo {title} {Optimal operation points for ultrafast, highly coherent ge hole spin-orbit qubits},\ }\href {https://doi.org/10.1038/s41534-021-00386-2} {\bibfield  {journal} {\bibinfo  {journal} {npj Quantum Inf.}\ }\textbf {\bibinfo {volume} {7}},\ \bibinfo {pages} {54} (\bibinfo {year} {2021})}\BibitemShut {NoStop}%
\bibitem [{\citenamefont {Bosco}\ \emph {et~al.}(2021)\citenamefont {Bosco}, \citenamefont {Benito}, \citenamefont {Adelsberger},\ and\ \citenamefont {Loss}}]{Bosco2021}%
  \BibitemOpen
  \bibfield  {author} {\bibinfo {author} {\bibfnamefont {S.}~\bibnamefont {Bosco}}, \bibinfo {author} {\bibfnamefont {M.}~\bibnamefont {Benito}}, \bibinfo {author} {\bibfnamefont {C.}~\bibnamefont {Adelsberger}},\ and\ \bibinfo {author} {\bibfnamefont {D.}~\bibnamefont {Loss}},\ }\bibfield  {title} {\bibinfo {title} {Squeezed hole spin qubits in ge quantum dots with ultrafast gates at low power},\ }\href {https://doi.org/10.1103/physrevb.104.115425} {\bibfield  {journal} {\bibinfo  {journal} {Phys. Rev. B}\ }\textbf {\bibinfo {volume} {104}},\ \bibinfo {pages} {115425} (\bibinfo {year} {2021})}\BibitemShut {NoStop}%
\bibitem [{\citenamefont {Wang}\ \emph {et~al.}(2024{\natexlab{a}})\citenamefont {Wang}, \citenamefont {Ercan}, \citenamefont {Gyure}, \citenamefont {Scappucci}, \citenamefont {Veldhorst},\ and\ \citenamefont {Rimbach-Russ}}]{Wang2}%
  \BibitemOpen
  \bibfield  {author} {\bibinfo {author} {\bibfnamefont {C.-A.}\ \bibnamefont {Wang}}, \bibinfo {author} {\bibfnamefont {H.~E.}\ \bibnamefont {Ercan}}, \bibinfo {author} {\bibfnamefont {M.~F.}\ \bibnamefont {Gyure}}, \bibinfo {author} {\bibfnamefont {G.}~\bibnamefont {Scappucci}}, \bibinfo {author} {\bibfnamefont {M.}~\bibnamefont {Veldhorst}},\ and\ \bibinfo {author} {\bibfnamefont {M.}~\bibnamefont {Rimbach-Russ}},\ }\bibfield  {title} {\bibinfo {title} {Modeling of planar germanium hole qubits in electric and magnetic fields},\ }\href {https://doi.org/10.1038/s41534-024-00897-8} {\bibfield  {journal} {\bibinfo  {journal} {npj Quantum Inf.}\ }\textbf {\bibinfo {volume} {10}},\ \bibinfo {pages} {102} (\bibinfo {year} {2024}{\natexlab{a}})}\BibitemShut {NoStop}%
\bibitem [{\citenamefont {Zajac}\ \emph {et~al.}(2016{\natexlab{b}})\citenamefont {Zajac}, \citenamefont {Hazard}, \citenamefont {Mi}, \citenamefont {Nielsen},\ and\ \citenamefont {Petta}}]{Zajac2016}%
  \BibitemOpen
  \bibfield  {author} {\bibinfo {author} {\bibfnamefont {D.}~\bibnamefont {Zajac}}, \bibinfo {author} {\bibfnamefont {T.}~\bibnamefont {Hazard}}, \bibinfo {author} {\bibfnamefont {X.}~\bibnamefont {Mi}}, \bibinfo {author} {\bibfnamefont {E.}~\bibnamefont {Nielsen}},\ and\ \bibinfo {author} {\bibfnamefont {J.}~\bibnamefont {Petta}},\ }\bibfield  {title} {\bibinfo {title} {Scalable gate architecture for a one-dimensional array of semiconductor spin qubits},\ }\href {https://doi.org/10.1103/physrevapplied.6.054013} {\bibfield  {journal} {\bibinfo  {journal} {Phys. Rev. Appl.}\ }\textbf {\bibinfo {volume} {6}},\ \bibinfo {pages} {054013} (\bibinfo {year} {2016}{\natexlab{b}})}\BibitemShut {NoStop}%
\bibitem [{\citenamefont {Ferdous}\ \emph {et~al.}(2018)\citenamefont {Ferdous}, \citenamefont {Chan}, \citenamefont {Veldhorst}, \citenamefont {Hwang}, \citenamefont {Yang}, \citenamefont {Sahasrabudhe}, \citenamefont {Klimeck}, \citenamefont {Morello}, \citenamefont {Dzurak},\ and\ \citenamefont {Rahman}}]{Ferdous2018}%
  \BibitemOpen
  \bibfield  {author} {\bibinfo {author} {\bibfnamefont {R.}~\bibnamefont {Ferdous}}, \bibinfo {author} {\bibfnamefont {K.~W.}\ \bibnamefont {Chan}}, \bibinfo {author} {\bibfnamefont {M.}~\bibnamefont {Veldhorst}}, \bibinfo {author} {\bibfnamefont {J.~C.~C.}\ \bibnamefont {Hwang}}, \bibinfo {author} {\bibfnamefont {C.~H.}\ \bibnamefont {Yang}}, \bibinfo {author} {\bibfnamefont {H.}~\bibnamefont {Sahasrabudhe}}, \bibinfo {author} {\bibfnamefont {G.}~\bibnamefont {Klimeck}}, \bibinfo {author} {\bibfnamefont {A.}~\bibnamefont {Morello}}, \bibinfo {author} {\bibfnamefont {A.~S.}\ \bibnamefont {Dzurak}},\ and\ \bibinfo {author} {\bibfnamefont {R.}~\bibnamefont {Rahman}},\ }\bibfield  {title} {\bibinfo {title} {Interface-induced spin-orbit interaction in silicon quantum dots and prospects for scalability},\ }\href {https://doi.org/10.1103/physrevb.97.241401} {\bibfield  {journal} {\bibinfo  {journal} {Phys. Rev. B}\ }\textbf {\bibinfo {volume} {97}},\ \bibinfo {pages} {241401} (\bibinfo {year} {2018})}\BibitemShut
  {NoStop}%
\bibitem [{\citenamefont {Heinz}\ and\ \citenamefont {Burkard}(2021)}]{Irina}%
  \BibitemOpen
  \bibfield  {author} {\bibinfo {author} {\bibfnamefont {I.}~\bibnamefont {Heinz}}\ and\ \bibinfo {author} {\bibfnamefont {G.}~\bibnamefont {Burkard}},\ }\bibfield  {title} {\bibinfo {title} {Crosstalk analysis for single-qubit and two-qubit gates in spin qubit arrays},\ }\href {https://doi.org/10.1103/physrevb.104.045420} {\bibfield  {journal} {\bibinfo  {journal} {Phys. Rev. B}\ }\textbf {\bibinfo {volume} {104}},\ \bibinfo {pages} {045420} (\bibinfo {year} {2021})}\BibitemShut {NoStop}%
\bibitem [{\citenamefont {Heinz}\ and\ \citenamefont {Burkard}(2022)}]{Irina2}%
  \BibitemOpen
  \bibfield  {author} {\bibinfo {author} {\bibfnamefont {I.}~\bibnamefont {Heinz}}\ and\ \bibinfo {author} {\bibfnamefont {G.}~\bibnamefont {Burkard}},\ }\bibfield  {title} {\bibinfo {title} {Crosstalk analysis for simultaneously driven two-qubit gates in spin qubit arrays},\ }\href {https://doi.org/10.1103/physrevb.105.085414} {\bibfield  {journal} {\bibinfo  {journal} {Phys. Rev. B}\ }\textbf {\bibinfo {volume} {105}},\ \bibinfo {pages} {085414} (\bibinfo {year} {2022})}\BibitemShut {NoStop}%
\bibitem [{\citenamefont {Undseth}\ \emph {et~al.}(2023)\citenamefont {Undseth}, \citenamefont {Xue}, \citenamefont {Mehmandoost}, \citenamefont {Rimbach-Russ}, \citenamefont {Eendebak}, \citenamefont {Samkharadze}, \citenamefont {Sammak}, \citenamefont {Dobrovitski}, \citenamefont {Scappucci},\ and\ \citenamefont {Vandersypen}}]{Undseth}%
  \BibitemOpen
  \bibfield  {author} {\bibinfo {author} {\bibfnamefont {B.}~\bibnamefont {Undseth}}, \bibinfo {author} {\bibfnamefont {X.}~\bibnamefont {Xue}}, \bibinfo {author} {\bibfnamefont {M.}~\bibnamefont {Mehmandoost}}, \bibinfo {author} {\bibfnamefont {M.}~\bibnamefont {Rimbach-Russ}}, \bibinfo {author} {\bibfnamefont {P.~T.}\ \bibnamefont {Eendebak}}, \bibinfo {author} {\bibfnamefont {N.}~\bibnamefont {Samkharadze}}, \bibinfo {author} {\bibfnamefont {A.}~\bibnamefont {Sammak}}, \bibinfo {author} {\bibfnamefont {V.~V.}\ \bibnamefont {Dobrovitski}}, \bibinfo {author} {\bibfnamefont {G.}~\bibnamefont {Scappucci}},\ and\ \bibinfo {author} {\bibfnamefont {L.~M.}\ \bibnamefont {Vandersypen}},\ }\bibfield  {title} {\bibinfo {title} {Nonlinear response and crosstalk of electrically driven silicon spin qubits},\ }\href {https://doi.org/10.1103/physrevapplied.19.044078} {\bibfield  {journal} {\bibinfo  {journal} {Phys. Rev. Appl.}\ }\textbf {\bibinfo {volume} {19}},\ \bibinfo {pages} {044078} (\bibinfo {year}
  {2023})}\BibitemShut {NoStop}%
\bibitem [{\citenamefont {Geyer}\ \emph {et~al.}(2024)\citenamefont {Geyer}, \citenamefont {Hetényi}, \citenamefont {Bosco}, \citenamefont {Camenzind}, \citenamefont {Eggli}, \citenamefont {Fuhrer}, \citenamefont {Loss}, \citenamefont {Warburton}, \citenamefont {Zumb\"{u}hl},\ and\ \citenamefont {Kuhlmann}}]{Geyer}%
  \BibitemOpen
  \bibfield  {author} {\bibinfo {author} {\bibfnamefont {S.}~\bibnamefont {Geyer}}, \bibinfo {author} {\bibfnamefont {B.}~\bibnamefont {Hetényi}}, \bibinfo {author} {\bibfnamefont {S.}~\bibnamefont {Bosco}}, \bibinfo {author} {\bibfnamefont {L.~C.}\ \bibnamefont {Camenzind}}, \bibinfo {author} {\bibfnamefont {R.~S.}\ \bibnamefont {Eggli}}, \bibinfo {author} {\bibfnamefont {A.}~\bibnamefont {Fuhrer}}, \bibinfo {author} {\bibfnamefont {D.}~\bibnamefont {Loss}}, \bibinfo {author} {\bibfnamefont {R.~J.}\ \bibnamefont {Warburton}}, \bibinfo {author} {\bibfnamefont {D.~M.}\ \bibnamefont {Zumb\"{u}hl}},\ and\ \bibinfo {author} {\bibfnamefont {A.~V.}\ \bibnamefont {Kuhlmann}},\ }\bibfield  {title} {\bibinfo {title} {Anisotropic exchange interaction of two hole-spin qubits},\ }\href {https://doi.org/10.1038/s41567-024-02481-5} {\bibfield  {journal} {\bibinfo  {journal} {Nat. Phys.}\ }\textbf {\bibinfo {volume} {20}},\ \bibinfo {pages} {1152–1157} (\bibinfo {year} {2024})}\BibitemShut {NoStop}%
\bibitem [{\citenamefont {Sarkar}\ \emph {et~al.}(2023)\citenamefont {Sarkar}, \citenamefont {Wang}, \citenamefont {Rendell}, \citenamefont {Hendrickx}, \citenamefont {Veldhorst}, \citenamefont {Scappucci}, \citenamefont {Khalifa}, \citenamefont {Salfi}, \citenamefont {Saraiva}, \citenamefont {Dzurak}, \citenamefont {Hamilton},\ and\ \citenamefont {Culcer}}]{Sarkar2023}%
  \BibitemOpen
  \bibfield  {author} {\bibinfo {author} {\bibfnamefont {A.}~\bibnamefont {Sarkar}}, \bibinfo {author} {\bibfnamefont {Z.}~\bibnamefont {Wang}}, \bibinfo {author} {\bibfnamefont {M.}~\bibnamefont {Rendell}}, \bibinfo {author} {\bibfnamefont {N.~W.}\ \bibnamefont {Hendrickx}}, \bibinfo {author} {\bibfnamefont {M.}~\bibnamefont {Veldhorst}}, \bibinfo {author} {\bibfnamefont {G.}~\bibnamefont {Scappucci}}, \bibinfo {author} {\bibfnamefont {M.}~\bibnamefont {Khalifa}}, \bibinfo {author} {\bibfnamefont {J.}~\bibnamefont {Salfi}}, \bibinfo {author} {\bibfnamefont {A.}~\bibnamefont {Saraiva}}, \bibinfo {author} {\bibfnamefont {A.~S.}\ \bibnamefont {Dzurak}}, \bibinfo {author} {\bibfnamefont {A.~R.}\ \bibnamefont {Hamilton}},\ and\ \bibinfo {author} {\bibfnamefont {D.}~\bibnamefont {Culcer}},\ }\bibfield  {title} {\bibinfo {title} {Electrical operation of planar ge hole spin qubits in an in-plane magnetic field},\ }\href {https://doi.org/10.1103/physrevb.108.245301} {\bibfield  {journal} {\bibinfo  {journal} {Phys.
  Rev. B}\ }\textbf {\bibinfo {volume} {108}},\ \bibinfo {pages} {245301} (\bibinfo {year} {2023})}\BibitemShut {NoStop}%
\bibitem [{\citenamefont {Cifuentes}\ \emph {et~al.}(2024)\citenamefont {Cifuentes}, \citenamefont {Tanttu}, \citenamefont {Steinacker}, \citenamefont {Serrano}, \citenamefont {Hansen}, \citenamefont {Slack-Smith}, \citenamefont {Gilbert}, \citenamefont {Huang}, \citenamefont {Vahapoglu}, \citenamefont {Leon}, \citenamefont {Stuyck}, \citenamefont {Itoh}, \citenamefont {Abrosimov}, \citenamefont {Pohl}, \citenamefont {Thewalt}, \citenamefont {Laucht}, \citenamefont {Yang}, \citenamefont {Escott}, \citenamefont {Hudson}, \citenamefont {Lim}, \citenamefont {Rahman}, \citenamefont {Dzurak},\ and\ \citenamefont {Saraiva}}]{Cifuentes}%
  \BibitemOpen
  \bibfield  {author} {\bibinfo {author} {\bibfnamefont {J.~D.}\ \bibnamefont {Cifuentes}}, \bibinfo {author} {\bibfnamefont {T.}~\bibnamefont {Tanttu}}, \bibinfo {author} {\bibfnamefont {P.}~\bibnamefont {Steinacker}}, \bibinfo {author} {\bibfnamefont {S.}~\bibnamefont {Serrano}}, \bibinfo {author} {\bibfnamefont {I.}~\bibnamefont {Hansen}}, \bibinfo {author} {\bibfnamefont {J.~P.}\ \bibnamefont {Slack-Smith}}, \bibinfo {author} {\bibfnamefont {W.}~\bibnamefont {Gilbert}}, \bibinfo {author} {\bibfnamefont {J.~Y.}\ \bibnamefont {Huang}}, \bibinfo {author} {\bibfnamefont {E.}~\bibnamefont {Vahapoglu}}, \bibinfo {author} {\bibfnamefont {R.~C.~C.}\ \bibnamefont {Leon}}, \bibinfo {author} {\bibfnamefont {N.~D.}\ \bibnamefont {Stuyck}}, \bibinfo {author} {\bibfnamefont {K.}~\bibnamefont {Itoh}}, \bibinfo {author} {\bibfnamefont {N.}~\bibnamefont {Abrosimov}}, \bibinfo {author} {\bibfnamefont {H.-J.}\ \bibnamefont {Pohl}}, \bibinfo {author} {\bibfnamefont {M.}~\bibnamefont {Thewalt}}, \bibinfo {author}
  {\bibfnamefont {A.}~\bibnamefont {Laucht}}, \bibinfo {author} {\bibfnamefont {C.~H.}\ \bibnamefont {Yang}}, \bibinfo {author} {\bibfnamefont {C.~C.}\ \bibnamefont {Escott}}, \bibinfo {author} {\bibfnamefont {F.~E.}\ \bibnamefont {Hudson}}, \bibinfo {author} {\bibfnamefont {W.~H.}\ \bibnamefont {Lim}}, \bibinfo {author} {\bibfnamefont {R.}~\bibnamefont {Rahman}}, \bibinfo {author} {\bibfnamefont {A.~S.}\ \bibnamefont {Dzurak}},\ and\ \bibinfo {author} {\bibfnamefont {A.}~\bibnamefont {Saraiva}},\ }\bibfield  {title} {\bibinfo {title} {Impact of electrostatic crosstalk on spin qubits in dense cmos quantum dot arrays},\ }\href {https://doi.org/10.1103/physrevb.110.125414} {\bibfield  {journal} {\bibinfo  {journal} {Phys. Rev. B}\ }\textbf {\bibinfo {volume} {110}},\ \bibinfo {pages} {125414} (\bibinfo {year} {2024})}\BibitemShut {NoStop}%
\bibitem [{\citenamefont {Wang}\ \emph {et~al.}(2024{\natexlab{b}})\citenamefont {Wang}, \citenamefont {John}, \citenamefont {Tidjani}, \citenamefont {Yu}, \citenamefont {Ivlev}, \citenamefont {Déprez}, \citenamefont {van Riggelen-Doelman}, \citenamefont {Woods}, \citenamefont {Hendrickx}, \citenamefont {Lawrie}, \citenamefont {Stehouwer}, \citenamefont {Oosterhout}, \citenamefont {Sammak}, \citenamefont {Friesen}, \citenamefont {Scappucci}, \citenamefont {de~Snoo}, \citenamefont {Rimbach-Russ}, \citenamefont {Borsoi},\ and\ \citenamefont {Veldhorst}}]{Wang2024}%
  \BibitemOpen
  \bibfield  {author} {\bibinfo {author} {\bibfnamefont {C.-A.}\ \bibnamefont {Wang}}, \bibinfo {author} {\bibfnamefont {V.}~\bibnamefont {John}}, \bibinfo {author} {\bibfnamefont {H.}~\bibnamefont {Tidjani}}, \bibinfo {author} {\bibfnamefont {C.~X.}\ \bibnamefont {Yu}}, \bibinfo {author} {\bibfnamefont {A.~S.}\ \bibnamefont {Ivlev}}, \bibinfo {author} {\bibfnamefont {C.}~\bibnamefont {Déprez}}, \bibinfo {author} {\bibfnamefont {F.}~\bibnamefont {van Riggelen-Doelman}}, \bibinfo {author} {\bibfnamefont {B.~D.}\ \bibnamefont {Woods}}, \bibinfo {author} {\bibfnamefont {N.~W.}\ \bibnamefont {Hendrickx}}, \bibinfo {author} {\bibfnamefont {W.~I.~L.}\ \bibnamefont {Lawrie}}, \bibinfo {author} {\bibfnamefont {L.~E.~A.}\ \bibnamefont {Stehouwer}}, \bibinfo {author} {\bibfnamefont {S.~D.}\ \bibnamefont {Oosterhout}}, \bibinfo {author} {\bibfnamefont {A.}~\bibnamefont {Sammak}}, \bibinfo {author} {\bibfnamefont {M.}~\bibnamefont {Friesen}}, \bibinfo {author} {\bibfnamefont {G.}~\bibnamefont {Scappucci}}, \bibinfo
  {author} {\bibfnamefont {S.~L.}\ \bibnamefont {de~Snoo}}, \bibinfo {author} {\bibfnamefont {M.}~\bibnamefont {Rimbach-Russ}}, \bibinfo {author} {\bibfnamefont {F.}~\bibnamefont {Borsoi}},\ and\ \bibinfo {author} {\bibfnamefont {M.}~\bibnamefont {Veldhorst}},\ }\bibfield  {title} {\bibinfo {title} {Operating semiconductor quantum processors with hopping spins},\ }\href {https://doi.org/10.1126/science.ado5915} {\bibfield  {journal} {\bibinfo  {journal} {Science}\ }\textbf {\bibinfo {volume} {385}},\ \bibinfo {pages} {447–452} (\bibinfo {year} {2024}{\natexlab{b}})}\BibitemShut {NoStop}%
\bibitem [{\citenamefont {Burkard}\ \emph {et~al.}(2023)\citenamefont {Burkard}, \citenamefont {Ladd}, \citenamefont {Pan}, \citenamefont {Nichol},\ and\ \citenamefont {Petta}}]{Burkard2}%
  \BibitemOpen
  \bibfield  {author} {\bibinfo {author} {\bibfnamefont {G.}~\bibnamefont {Burkard}}, \bibinfo {author} {\bibfnamefont {T.~D.}\ \bibnamefont {Ladd}}, \bibinfo {author} {\bibfnamefont {A.}~\bibnamefont {Pan}}, \bibinfo {author} {\bibfnamefont {J.~M.}\ \bibnamefont {Nichol}},\ and\ \bibinfo {author} {\bibfnamefont {J.~R.}\ \bibnamefont {Petta}},\ }\bibfield  {title} {\bibinfo {title} {Semiconductor spin qubits},\ }\href {https://doi.org/10.1103/revmodphys.95.025003} {\bibfield  {journal} {\bibinfo  {journal} {Rev. Mod. Phys.}\ }\textbf {\bibinfo {volume} {95}},\ \bibinfo {pages} {025003} (\bibinfo {year} {2023})}\BibitemShut {NoStop}%
\bibitem [{\citenamefont {Frank}\ \emph {et~al.}(2020)\citenamefont {Frank}, \citenamefont {Scher\"{u}bl}, \citenamefont {Csonka}, \citenamefont {Zaránd},\ and\ \citenamefont {Pályi}}]{Frank}%
  \BibitemOpen
  \bibfield  {author} {\bibinfo {author} {\bibfnamefont {G.}~\bibnamefont {Frank}}, \bibinfo {author} {\bibfnamefont {Z.}~\bibnamefont {Scher\"{u}bl}}, \bibinfo {author} {\bibfnamefont {S.}~\bibnamefont {Csonka}}, \bibinfo {author} {\bibfnamefont {G.}~\bibnamefont {Zaránd}},\ and\ \bibinfo {author} {\bibfnamefont {A.}~\bibnamefont {Pályi}},\ }\bibfield  {title} {\bibinfo {title} {Magnetic degeneracy points in interacting two-spin systems: Geometrical patterns, topological charge distributions, and their stability},\ }\href {https://doi.org/10.1103/physrevb.101.245409} {\bibfield  {journal} {\bibinfo  {journal} {Phys. Rev. B}\ }\textbf {\bibinfo {volume} {101}},\ \bibinfo {pages} {101} (\bibinfo {year} {2020})}\BibitemShut {NoStop}%
\bibitem [{\citenamefont {Pedersen}\ \emph {et~al.}(2007)\citenamefont {Pedersen}, \citenamefont {Møller},\ and\ \citenamefont {Mølmer}}]{Pedersen}%
  \BibitemOpen
  \bibfield  {author} {\bibinfo {author} {\bibfnamefont {L.~H.}\ \bibnamefont {Pedersen}}, \bibinfo {author} {\bibfnamefont {N.~M.}\ \bibnamefont {Møller}},\ and\ \bibinfo {author} {\bibfnamefont {K.}~\bibnamefont {Mølmer}},\ }\bibfield  {title} {\bibinfo {title} {Fidelity of quantum operations},\ }\href {https://doi.org/10.1016/j.physleta.2007.02.069} {\bibfield  {journal} {\bibinfo  {journal} {Phys. Lett. A}\ }\textbf {\bibinfo {volume} {367}},\ \bibinfo {pages} {47–51} (\bibinfo {year} {2007})}\BibitemShut {NoStop}%
\bibitem [{\citenamefont {Kelly}\ \emph {et~al.}(2023)\citenamefont {Kelly}, \citenamefont {Orekhov}, \citenamefont {Hendrickx}, \citenamefont {Mergenthaler}, \citenamefont {Schupp}, \citenamefont {Paredes}, \citenamefont {Eggli}, \citenamefont {Kuhlmann}, \citenamefont {Harvey-Collard}, \citenamefont {Fuhrer},\ and\ \citenamefont {Salis}}]{Kelly}%
  \BibitemOpen
  \bibfield  {author} {\bibinfo {author} {\bibfnamefont {E.~G.}\ \bibnamefont {Kelly}}, \bibinfo {author} {\bibfnamefont {A.}~\bibnamefont {Orekhov}}, \bibinfo {author} {\bibfnamefont {N.~W.}\ \bibnamefont {Hendrickx}}, \bibinfo {author} {\bibfnamefont {M.}~\bibnamefont {Mergenthaler}}, \bibinfo {author} {\bibfnamefont {F.~J.}\ \bibnamefont {Schupp}}, \bibinfo {author} {\bibfnamefont {S.}~\bibnamefont {Paredes}}, \bibinfo {author} {\bibfnamefont {R.~S.}\ \bibnamefont {Eggli}}, \bibinfo {author} {\bibfnamefont {A.~V.}\ \bibnamefont {Kuhlmann}}, \bibinfo {author} {\bibfnamefont {P.}~\bibnamefont {Harvey-Collard}}, \bibinfo {author} {\bibfnamefont {A.}~\bibnamefont {Fuhrer}},\ and\ \bibinfo {author} {\bibfnamefont {G.}~\bibnamefont {Salis}},\ }\bibfield  {title} {\bibinfo {title} {Capacitive crosstalk in gate-based dispersive sensing of spin qubits},\ }\href {https://doi.org/10.1063/5.0177857} {\bibfield  {journal} {\bibinfo  {journal} {Appl. Phys. Lett.}\ }\textbf {\bibinfo {volume} {123}},\ \bibinfo {pages}
  {0177857} (\bibinfo {year} {2023})}\BibitemShut {NoStop}%
\bibitem [{\citenamefont {Cayao}\ \emph {et~al.}(2020)\citenamefont {Cayao}, \citenamefont {Benito},\ and\ \citenamefont {Burkard}}]{Cayao}%
  \BibitemOpen
  \bibfield  {author} {\bibinfo {author} {\bibfnamefont {J.}~\bibnamefont {Cayao}}, \bibinfo {author} {\bibfnamefont {M.}~\bibnamefont {Benito}},\ and\ \bibinfo {author} {\bibfnamefont {G.}~\bibnamefont {Burkard}},\ }\bibfield  {title} {\bibinfo {title} {Programable two-qubit gates in capacitively coupled flopping-mode spin qubits},\ }\href {https://doi.org/10.1103/physrevb.101.195438} {\bibfield  {journal} {\bibinfo  {journal} {Phys. Rev. B}\ }\textbf {\bibinfo {volume} {101}},\ \bibinfo {pages} {195438} (\bibinfo {year} {2020})}\BibitemShut {NoStop}%
\bibitem [{\citenamefont {Neyens}\ \emph {et~al.}(2019)\citenamefont {Neyens}, \citenamefont {MacQuarrie}, \citenamefont {Dodson}, \citenamefont {Corrigan}, \citenamefont {Holman}, \citenamefont {Thorgrimsson}, \citenamefont {Palma}, \citenamefont {McJunkin}, \citenamefont {Edge}, \citenamefont {Friesen}, \citenamefont {Coppersmith},\ and\ \citenamefont {Eriksson}}]{Neyens2019}%
  \BibitemOpen
  \bibfield  {author} {\bibinfo {author} {\bibfnamefont {S.~F.}\ \bibnamefont {Neyens}}, \bibinfo {author} {\bibfnamefont {E.}~\bibnamefont {MacQuarrie}}, \bibinfo {author} {\bibfnamefont {J.}~\bibnamefont {Dodson}}, \bibinfo {author} {\bibfnamefont {J.}~\bibnamefont {Corrigan}}, \bibinfo {author} {\bibfnamefont {N.}~\bibnamefont {Holman}}, \bibinfo {author} {\bibfnamefont {B.}~\bibnamefont {Thorgrimsson}}, \bibinfo {author} {\bibfnamefont {M.}~\bibnamefont {Palma}}, \bibinfo {author} {\bibfnamefont {T.}~\bibnamefont {McJunkin}}, \bibinfo {author} {\bibfnamefont {L.}~\bibnamefont {Edge}}, \bibinfo {author} {\bibfnamefont {M.}~\bibnamefont {Friesen}}, \bibinfo {author} {\bibfnamefont {S.}~\bibnamefont {Coppersmith}},\ and\ \bibinfo {author} {\bibfnamefont {M.}~\bibnamefont {Eriksson}},\ }\bibfield  {title} {\bibinfo {title} {Measurements of capacitive coupling within a quadruple-quantum-dot array},\ }\href {https://doi.org/10.1103/physrevapplied.12.064049} {\bibfield  {journal} {\bibinfo  {journal} {Phys. Rev.
  Appl.}\ }\textbf {\bibinfo {volume} {12}},\ \bibinfo {pages} {064049} (\bibinfo {year} {2019})}\BibitemShut {NoStop}%
\end{thebibliography}%


\end{document}